\documentclass[journal]{IEEEtran}
\ifCLASSINFOpdf
\else
\fi
\usepackage[cmex10]{amsmath}
\usepackage{ amssymb }
\usepackage[pdftex]{graphicx}
\usepackage{float}
\usepackage{multicol}
\usepackage{lipsum}
\usepackage{ifpdf}
\usepackage{multirow}
\usepackage{epstopdf}
\usepackage{color}
\usepackage{cite}
\usepackage{graphicx}
\usepackage{soul}
\usepackage{balance}
\usepackage{threeparttable}
\usepackage{wasysym}
\usepackage{url}
\usepackage[normalsize]{subfigure}
\usepackage{enumitem}
\usepackage{enumitem}
\usepackage{dblfloatfix}    % To enable figures at the bottom of page
\usepackage{afterpage}
\usepackage{changepage} 
\usepackage[hidelinks]{hyperref}
\usepackage{accents}
\usepackage{algorithmicx}
\usepackage{algpseudocode}

\newcommand{\ubar}[1]{\underaccent{\bar}{#1}}

\usepackage{fancyhdr}

{}
\begin{document}

\IEEEpeerreviewmaketitle
%
% paper title
% Titles are generally capitalized except for words such as a, an, and, as,
% at, but, by, for, in, nor, of, on, or, the, to and up, which are usually
% not capitalized unless they are the first or last word of the title.
% Linebreaks \\ can be used within to get better formatting as desired.
% Do not put math or special symbols in the title.
\title{Loss Minimization with Optimal Power Dispatch in Multi-Frequency HVac Power Systems}
%
%
% author names and IEEE memberships
% note positions of commas and nonbreaking spaces ( ~ ) LaTeX will not break
% a structure at a ~ so this keeps an author's name from being broken across
% two lines.
% use \thanks{} to gain access to the first footnote area
% a separate \thanks must be used for each paragraph as LaTeX2e's \thanks
% was not built to handle multiple paragraphs
%

\author{Quan~Nguyen,~\IEEEmembership{Student Member,~IEEE,}
		Keng-Weng Lao, \IEEEmembership{Member,~IEEE,}
		Phuong Vu,		
        and~Surya~Santoso,~\IEEEmembership{Fellow,~IEEE}% <-this % stops a space
%\thanks{The authors are with the Department of Electrical and Computer Engineering, University of Texas at Austin (e-mail: quan.nguyenhuy@utexas.edu; johnnylao@utexas.edu; ssantoso@mail.utexas.edu) and Hanoi University of Science and Technology (e-mail:phuong.vuhoang@hust.edu.vn).}% <-this % stops a space
}

% make the title area
\maketitle

\thispagestyle{fancy}

% As a general rule, do not put math, special symbols or citations
% in the abstract or keywords.
\begin{abstract}
Low-frequency high voltage ac transmission scheme has recently been proposed as an alternative approach for bulk power transmission. This paper proposes a multi-period optimal power flow (OPF) for a multi-frequency HVac transmission system that interconnects both conventional 50/60-Hz and low-frequency grids using back-to-back converters with a centralized control scheme. The OPF objective is to minimize system losses by determining the optimal dispatch for generators, shunt capacitors, and converters. The OPF constraints include the operational constraints of all HVac grid and converter stations. The resulting mixed-integer nonlinear programing problem is solved using a proposed framework based on the predictor-corrector primal-dual interior-point method. The proposed OPF formulation and solution approach are verified using a multi-frequency HVac transmission system that is modified from the IEEE 57-bus system. The results with the optimal dispatch from the proposed method during a simulated day show a significant loss reduction and an improved voltage regulation compared to those when an arbitrary dispatch is chosen.
\end{abstract}

% Note that keywords are not normally used for peerreview papers.

% For peer review papers, you can put extra information on the cover
% page as needed:
% \ifCLASSOPTIONpeerreview
% \begin{center} \bfseries EDICS Category: 3-BBND \end{center}
% \fi
%
% For peerreview papers, this IEEEtran command inserts a page break and
% creates the second title. It will be ignored for other modes.
\IEEEpeerreviewmaketitle

\section{Nomenclature}
Subscripts $*$ for power grid names:
\vspace{0.1cm}

\begin{tabular}{p{1.8cm} p{5.8cm}}
	\textit{$s$}          & Conventional 50/60-Hz HVac grid; \\
	\textit{$l$}          & LF-HVac grid; \\
\end{tabular}\\

Sets:
\vspace{0.1cm}

\begin{tabular}{p{1.8cm} p{5.8cm}}
	\textit{$\mathcal{N}_*$}          		& set of buses in a grid; \\ 
	\textit{$\mathcal{D}_{*}$}      		& set of transmission lines;    \\	
	\textit{$\mathcal{G}_*$}          		& set of buses connected to generators; \\	
	\textit{$\mathcal{L_*}$}          		& set of non-voltage-controlled buses; \\ 
	\textit{$\mathcal{C_*}$}      	  		& set of buses connected to converters;\\ 	
	\textit{$\mathcal{V}_*$}          		& set of buses connected to converters operating in voltage-controlled mode; \\	
	\textit{$\mathcal{N}_{*}^{sh}$}    		& set of buses with shunt capacitors; 	\\
	\textit{$\mathcal{Q}_{*, k}^{sh}$} 		& set of discrete dispatch of the shunt capacitor at bus $k$; 	
\end{tabular}\\

Parameters:
\vspace{0.1cm}

\begin{tabular}{p{1.8cm} p{5.8cm}}
	\textit{$P_{s}^{load}, Q_{s}^{load}$}   & loads at buses in HVac grid $s$;\\	
	\textit{$g_{*}, b_{*}$}   & line series conductance and susceptance;\\	
	\textit{$V_{*}^{o}$}   					& reference voltage magnitudes at voltage-controlled buses;\\	
	\textit{${\ubar{V}}, {\bar{V}}$}		& lower and upper load voltage limits;\\	
	\textit{${\bar{I}_{*}}$}				& maximum line current;\\		
	\textit{${\bar{P}_{*}}$}				& maximum line active power;\\	
	\textit{$\boldsymbol{Y}_{*}$, $\boldsymbol{G}_{*}$, $\boldsymbol{B}_{*}$}			& admittance, conductance, and susceptance matrices;\\	
	\textit{$\alpha_1$, $\alpha_2$} 		& weighting coefficients;\\
\end{tabular}\\

Variables:
\vspace{0.1cm}

\begin{tabular}{p{1.8cm} p{5.8cm}}
	\textit{$e_*, f_*$}    					& real and imaginary parts of voltages;\\
	\textit{$P_*, Q_*$}         			& injected power into grid from buses;	
\end{tabular}

\begin{tabular}{p{1.8cm} p{5.8cm}}
	\textit{$P_*^{gen}, Q_*^{gen}$}         & generator dispatch;\\
	\textit{$P_*^{conv}, Q_*^{conv}$}       & power from/to BTB converters;	\\	
	\textit{$Q_*^{sh}$}        			    & injected reactive power into grid from shunt capacitors at 1 pu;
\end{tabular}

\section{Introduction}
Recent developments of flexible alternating current transmission system (FACTS) and control techniques greatly affect power system infrastructure and operational characteristics. Compared to the conventional high-voltage alternating current (HVac) transmission systems, converter-based high-voltage direct current (HVdc) systems have shown unique benefits in transmitting bulk electric power from remote generation stations to load centers. HVdc systems offer unrestricted point-to-point long-distance transmission capability, reduced conductor loss, narrower right of way, and the ability to interconnect ac systems with different operating frequencies. Such advantages, however, cannot justify for the lack of reliable system protection devices, which impedes the feasibility of flexible and reliable multi-point interconnection capability as in HVac systems.

Recently, low frequency HVac (LF-HVac) has been proposed as an alternate solution for HVac and HVdc systems. LF-HVac systems combine the advantages of the two existing technologies, such as high power-carrying capability over long distance, straightforward ac protection system, and the use of multi-terminal networks \cite{Funaki_1,Fischer_1,PSERC_1}. In \cite{Tuan_1}, it is shown that a significant reduction in reactance at low frequency benefits LF-HVac transmission systems in terms of low-load and full-load voltage profiles. Voltage stability and system dynamic response following a disturbance are also improved in LF-HVac systems compared to conventional HVac systems \cite{Tuan_2}\cite{Rosewater_1}. The possibility of implementing 16.7 Hz LF-HVac transmission for practical offshore wind farms in Europe is investigated in \cite{Tom_1} and \cite{Erlich_1}.

Similar to HVdc systems, an LF-HVac system requires power converters for connection to a conventional 50/60-Hz HVac system, which forms a multi-frequency power system. Besides frequency conversion, these converters offer additional control capability in the entire transmission system. Specifically, the power flow transfer between two buses in the LF-HVac and 50/60-Hz systems, i.e. the power sent to one bus from the other one, is controlled by the power set point of the converter between the two buses. In addition to the existing generators and shunt capacitors, the converters in an interconnected multi-frequency HVac transmission system can be considered as additional control resources to achieve an optimal operation with minimum system losses or generation cost. Unlike shunt capacitors, that reactive power from the converters are able to adapt to the fast dynamics of demands makes them superior in providing ancillary services such as voltage regulation. A generalized optimal power flow (OPF) model for multi-frequency HVac transmission systems incorporating converter model, control, and losses is thus needed to further optimize system operation subjected to operational constraints and demand variations.

Towards this end, several research has been done. The design of a grid-forming control for back-to-back (BTB) voltage-source converters (VSC) to maintain a stable offshore voltage for a long LF-HVac system is discussed in \cite{Ruddy_2}. In \cite{Quan_3}, the control and coordination of BTB converters to form an LF-HVac grid and connect to a standard 50/60-Hz HVac and HVdc grids are presented. A generalized unified power flow (PF) formulation in polar coordinates and solution for such a multi-frequency HVac - HVdc power system are also included. With the inclusion of BTB converters connecting two systems at different frequencies, PF can be solved using sequential or unified approaches \cite{Jef_2,Baradar_1,Quan_0}. The latter is more convenient to be implemented in solving OPF problems. To the best knowledge of the authors, however, no research has been done to solve OPF in multi-frequency HVac power systems.

In conventional 50/60-Hz HVac systems, OPF is characterized as a nonconvex mixed-integer nonlinear programing (MINLP) problem. Existing approaches to solve this OPF for conventional HVac systems include linearization, relaxation, and nonlinear programing. Linearization methods suffers poor solution accuracy in spite of the availability of reliable solvers. On the other hand, the solution obtained from relaxation methods might be unable to recover the solution of the original problem \cite{Farivar_2, Lavaei_1, Capitanescu_1}. The third approach addresses the exact nonlinear model of power system, so the accuracy and feasibility of the obtained solution is guaranteed. This approach, however, is non-polynomial hard, and it only guarantees local optimum in near real-time applications. 

One of the most effective methods for solving nonconvex NLP problems is the predictor-corrector primal-dual interior point method (PCPDIPM) \cite{Torres_1,Liu_1,Quan_2}. Several widely-used open-source NLP solvers such as IPOPT were developed based on this method \cite{IPOPT}. However, the derivative approximations used in these interior-point-based solver may seriously degrade the performance of the solver \cite{IPOPT, KNITRO}. In addition, their applications are limited to NLP and not applicable to MINLP problems. On the other hand, MINLP solvers such as BONMIN implement branching algorithms to efficiently handle discrete variables \cite{BONMIN}. However, these algorithms might take more time and iterations to converge than the interior point algorithm.

As an extension of the existing works on the emerging LF-HVac transmission technology, the main contributions of this paper are:
\begin{itemize}[leftmargin=*]
	\item A multi-period OPF formulation for the optimal operation of a multi-frequency HVac transmission system employing BTB converters with a centralized control scheme. The optimal dispatch of generators and BTB converters as well as capacitor switching pattern are determined to minimize the system losses, subjected to comprehensive operational constraints of ac grids and converter stations.
	\item An exact formulation of Hessian matrices corresponding to converter constraints that is required in a developed solution framework based on PCPDIPM. In addition, compared to \cite{Quan_3}, the OPF formulation is written in rectangular coordinates to take the advantage of the constant Hessian matrices of nodal power balance constraints. The derived Hessian matrices and the formulation in rectangular coordinates thus reduce the computational burden and convergence time for real-time applications. Since the formulated OPF is nonconvex, this approach only guarantees a local solution.
\end{itemize}  
The proposed OPF model and solution can be used for both operational and planning analysis of bulk power transmission systems from remote generation units to conventional 50/60-Hz HVac systems with an embedded LF-HVac system. This framework is also able to extend to include HVac - HVdc interconnections.

\section{System Structure and Centralized Control for BTB Conveters}\label{sec:Modeling}
\begin{figure}[t!]
	\centering				
	\includegraphics[width = 0.9\columnwidth] {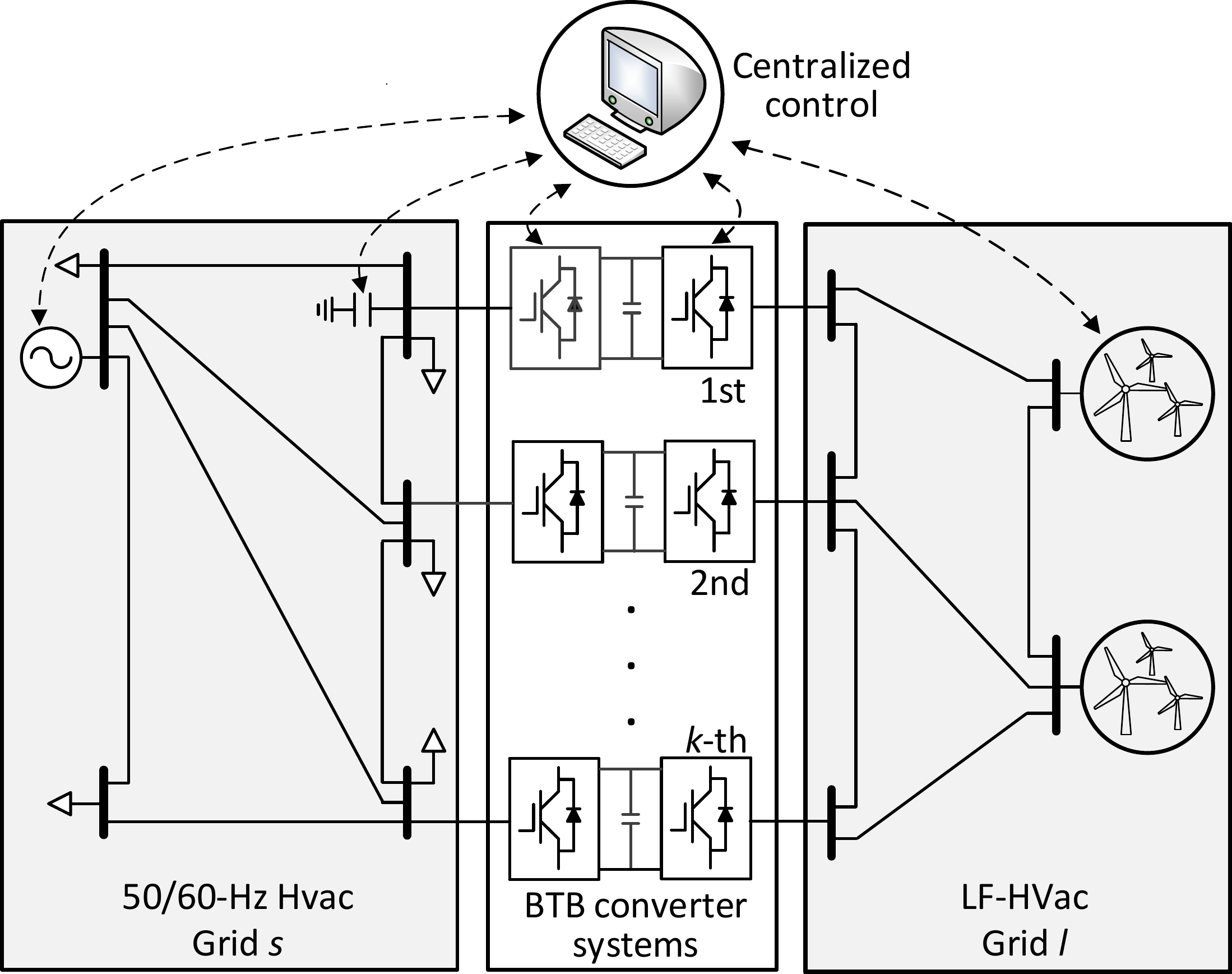}	
	\caption{An example of a multi-frequency power systems. The HVac grid $s$ and LF-HVac grid $l$ are connected to each other through BTB converters using centralized control. Loads are assumed to be located only in HVac grid $s$.}
	\label{fig:Back-to-Back_Converter}
\end{figure}

This section focuses on the modeling of BTB converters employing a centralized control in a generalized multi-frequency HVac system similar to the one shown in Fig. \ref{fig:Back-to-Back_Converter}.

\subsection{Steady-State Model of BTB Converter Stations}

\setcounter{figure}{1}
\begin{figure*}[t!]
	\centering
	\includegraphics[width = 1.8\columnwidth] {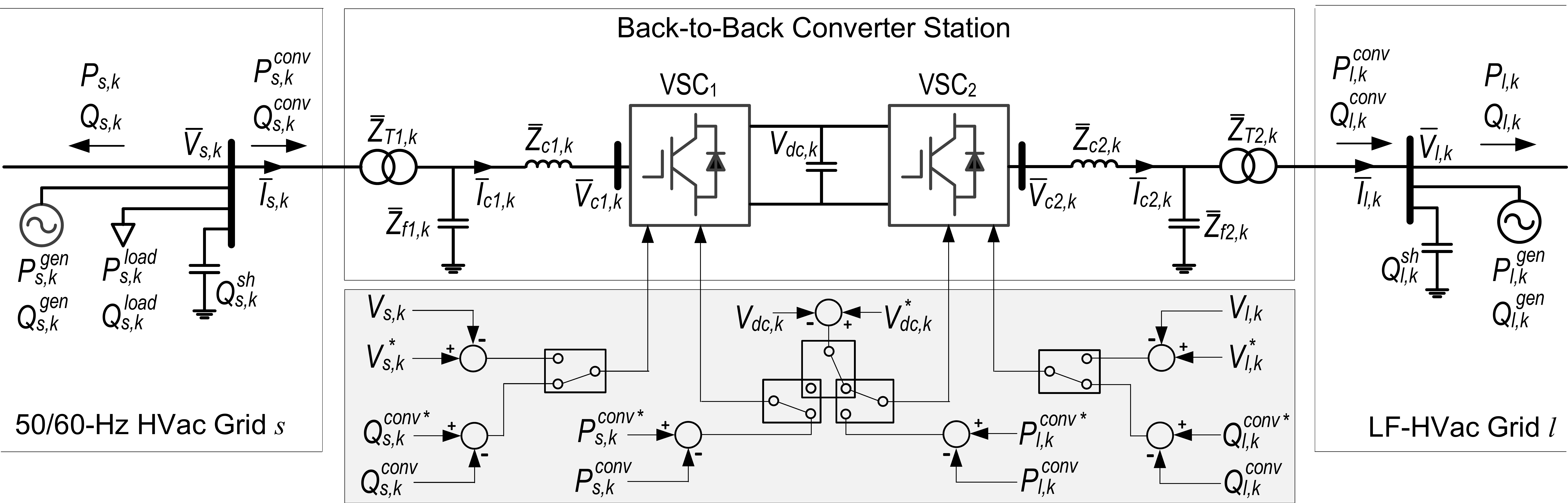}	
	\caption{A BTB converter is used to connect an LF-HVac grid to a 50/60-Hz HVac grid: the system configuration, the interface between the two grids, and the main control blocks. }
	\label{fig:Back-to-Back_Converter_Control}
\end{figure*}
Fig. \ref{fig:Back-to-Back_Converter_Control} shows the model and control of each BTB converter in Fig. \ref{fig:Back-to-Back_Converter}, which consists of two VSC denoted as $\textup{VSC}_\textup{1}$ and $\textup{VSC}_\textup{2}$. These converters have identical structures and electrical components such as a transformer, a shunt capacitor,  a phase reactor, and switching valves. These two converters share a common dc-link capacitor, which acts as an intermediate energy storage component and allows decoupled operation of converters $\textup{VSC}_\textup{1}$ and $\textup{VSC}_\textup{2}$. Each VSC converter can be thought of as a synchronous machine without inertia. Assuming converter voltages $\bar{V}_{c1}$ and $\bar{V}_{c2}$ contain no harmonic, each VSC converter is represented as a controllable voltage source behind an impedance \cite{Jef_2}. The impedances $\bar{Z}_{T1}$ and $\bar{Z}_{T2}$, $\bar{Z}_{f1}$ and $\bar{Z}_{f2}$, and $\bar{Z}_{c1}$ and $\bar{Z}_{c2}$ represent the transformer leakage impedances, shunt capacitor impedances, and the phase reactor impedances, respectively, at both side of the BTB converter.

\subsection{Operation Modes and Coordination of BTB Converters with Centralized Control in Multi-Frequency HVac Systems}
As described above, the operations of $\textup{VSC}_\textup{1}$ and $\textup{VSC}_\textup{2}$ are decoupled by keeping the dc-link capacitor voltage  $V_{dc,k}$ constant. At each side, the active and reactive power injected or withdrawn from the corresponding ac grid are also controlled independently \cite{Teodorescu_1,Li_1}. The real power control target for $P_{s,k}^{conv}$ and $P_{l,k}^{conv}$ can be changed to regulate the dc-link voltage $V_{dc,k}$. On the other hand, the reactive power control target for $Q_{s,k}^{conv}$ and $Q_{l,k}^{conv}$ can be switched to regulate the voltage magnitude $V_{s,k}$ and $V_{l,k}$ of the associated bus in HVac grid $s$ or LF-HVac grid $l$, respectively. 

When $n$ BTB converters are used as power flow controllers between HVac grid $s$ and LF-HVac grid $l$, as shown in Fig. \ref{fig:Back-to-Back_Converter}, the following operating modes are proposed to regulate the power, dc-link voltage, and/or ac voltage magnitude of the ac bus at each side of $n$-1 BTB converters:

\begin{itemize}[leftmargin=*]
	\item Converter $\textup{VSC}_\textup{1}$, which is connected to HVac grid $s$, is set to regulate active and reactive power (PQ mode) or active power and voltage magnitude at the $k^{th}$ bus in HVac grid $s$ (PV mode). This means that $P_{s,k}^{conv}$ and $Q_{s,k}^{conv}$ in PQ mode or $P_{s,k}^{conv}$ and $V_{s,k}$ in PV mode are the input for the OPF analysis.
	\item  Converter $\textup{VSC}_\textup{2}$, which is connected to LF-HVac grid $l$, is set to regulate the dc-link voltage and to control either reactive power (Q$\textup{V}_\textup{dc}$ mode) or the terminal voltage magnitude (V$\textup{V}_\textup{dc}$ mode). This implies that either $Q_{l,k}^{conv}$ in Q$\textup{V}_\textup{dc}$ mode or $V_{l,k}$ in V$\textup{V}_\textup{dc}$ mode is a known quantity while $P_{l,k}^{conv}$ is unknown and needs to be determined.
\end{itemize}

This approach does not apply to one particular BTB converter, which varies the unknown reactive power $Q_{l,sl}^{conv}$ at the $\textup{VSC}_\textup{2}$ side to control the voltage magnitude $V_{l,sl}$ at the slack bus of LF-HVac grid $l$. $\textup{VSC}_\textup{2}$ also regulates the active power $P_{l,sl}^{conv}$ flowing toward $\textup{VSC}_\textup{2}$ from the slack bus. However, the set point of $P_{l,sl}^{conv}$ is unknown since it depends on the losses in LF-HVac grid $l$. Since the phase angle reference is lost due to the intermediate dc stage in the ac/dc/ac conversion, the phase angle of the slack bus in LF-HVac grid $l$ is considered to be zero. On the other hand, $\textup{VSC}_\textup{1}$ side of this BTB converter controls the reactive power $Q_{s,sl}^{conv}$ and $\textup{V}_\textup{dc}$. The active power $P_{s,sl}^{conv}$ is unknown since it depends on the unknown active power $P_{l,sl}^{conv}$, on the Joule losses due to the real parts of $\bar{Z}_{T1}$, $\bar{Z}_{c1}$, $\bar{Z}_{T2}$, and $\bar{Z}_{c2}$, and on the switching losses in $\textup{VSC}_\textup{1}$ and $\textup{VSC}_\textup{2}$. Note that the ac bus connected to the $\textup{VSC}_\textup{1}$ side of this converter does not need to be the slack bus of HVac grid $s$.

\section{OPF Formulation in LF-HVAC Power Systems with BTB Converters}\label{sec:Formulation}
This section proposes an OPF formulation in the rectangular coordinates to determine the optimal dispatch of generators, shunt capacitors, and converters for minimizing system losses in a generalized multi-frequency system described in Section \ref{sec:Modeling}.

\subsection{Variables}
In rectangular coordinates, the state variable $\boldsymbol{x}$ includes the real and imaginary parts ($\boldsymbol{e}_{s}, \boldsymbol{f}_{s}$) of bus voltages in HVac grid $s$ and ($\boldsymbol{e}_{l}, \boldsymbol{f}_{l}$) in LF-HVac grid $l$. The decision variable $\boldsymbol{u}$ includes the dispatch of generators and shunt capacitors in HVac grid $s$ ($\boldsymbol{P}_{s}^{gen},{\boldsymbol{Q}_{s}^{gen}}, \boldsymbol{Q}_s^{sh}$)  and LF-HVac grid $l$ ($\boldsymbol{P}_{l}^{gen},{\boldsymbol{Q}_{l}^{gen}}, \boldsymbol{Q}_l^{sh}$). The dispatch at two sides of all BTB converters (${\boldsymbol{P}_{s}^{conv}}, {\boldsymbol{Q}_{s}^{conv}}, {\boldsymbol{P}_{l}^{conv}}, {\boldsymbol{Q}_{l}^{conv}}$) are the other components of $\boldsymbol{u}$. The combined variable vector of the proposed OPF problem is thus defined as follows:
\begin{align}
\label{eqn:Variables}
\nonumber
\boldsymbol{X} = [\boldsymbol{x} | \boldsymbol{u}] = [\boldsymbol{e}_{s}, \boldsymbol{f}_{s}, \boldsymbol{e}_{l}, &\boldsymbol{f}_{l} \hspace{0.1cm} | \hspace{0.1cm} {\boldsymbol{P}_{s}^{gen}}, {\boldsymbol{Q}_{s}^{gen}}, {\boldsymbol{Q}_{s}^{sh}},\\
{\boldsymbol{P}_{l}^{gen}}, {\boldsymbol{Q}_{l}^{gen}}, \boldsymbol{Q}_l^{sh}, & \hspace{0.1cm} {\boldsymbol{P}_{s}^{conv}}, {\boldsymbol{Q}_{s}^{conv}}, {\boldsymbol{P}_{l}^{conv}}, {\boldsymbol{Q}_{l}^{conv}}].
\end{align}

\subsection{Weighted-sum Objective Functions}
The objective of this OPF problem in this paper is to minimize system losses and shunt capacitor switching operations for a given operating horizon. Within a time step, the total measured load is assumed to be constant. Therefore, minimizing the system losses is equal to minimizing the sum of active power generated from all generators in HVac grid $s$ and LF-HVac grid $l$. The objective function of the optimal power flow is thus defined as follows: %The second component represents the sum of the power deviation at each generator from its schedule values, which is given as the solution of the UC \& ED problem. $\pi_{P}$ is a weighting coefficient corresponding to the power deviation.

\begin{align}
\label{eqn:Objectives}
\nonumber
f(\boldsymbol{X}) &= \alpha_1 \big[\sum_{k \in {\mathcal{G}_s}} \! P_{s,k}^{gen}+\sum_{k \in {\mathcal{G}_l}} \! P_{l,k}^{gen} \big] \\
 +  \alpha_2 & \big[\sum_{k \in \mathcal{N}_{s}^{sh}} \!\!\!{(Q_{s,k}^{sh} \!-\! Q_{s,k}^{sh^{pre} })}^{2} + \sum_{k \in \mathcal{N}_{l}^{sh}} \!\!\!{(Q_{l,k}^{sh} \!-\! Q_{l,k}^{sh^{pre} })}^{2} \big],
\end{align}
where $Q_{s,k}^{sh^{pre} }$ and $Q_{l,k}^{sh^{pre} }$ are the dispatch during the previous time step of the shunt capacitor at bus $k$ in HVac grid $s$ and LF-HVac grid $l$, respectively.

\subsection{Constraints of HVac Grid $s$}
The equality constraints $\boldsymbol{g}_{s}^{PQ}$ in HVac grid $s$, representing the active and reactive power balance at all buses, are given by:
\begin{align}
\label{eqn:AC_PF_eqn}	
\nonumber
g_{s,k}^{P}(\boldsymbol{X}) &= P_{s,k} - (P_{s, k}^{gen} - P_{s, k}^{load} - P_{s,k}^{conv}) = 0, \\
\nonumber
g_{s,k}^{Q}(\boldsymbol{X}) &= Q_{s,k} - (Q_{s, k}^{gen} - Q_{s, k}^{load} - Q_{s,k}^{conv}) \\
         &\hspace{1.08cm}- (e_{s,k}^2+f_{s,k}^2)Q_{s,k}^{sh} = 0, \hspace{.2cm} \forall k \in \mathcal{N}_{s}.
\end{align}
The power $P_{s,k}$ and $Q_{s,kk}$ injected into HVac grid $s$ at bus $k$ in (\ref{eqn:AC_PF_eqn}) are as determined follows:	
\begin{align}
\label{eqn:AC_injected_power_eqn}
\nonumber
P_{s,k} = \boldsymbol{G}_{s,k:} (e_{s,k}\boldsymbol{e}_{s}  +  f_{s,k}\boldsymbol{f}_{s})  +  \boldsymbol{B}_{s,k:} &(f_{s,k}\boldsymbol{e}_{s} - e_{s,k}\boldsymbol{f}_{s}),\\
\nonumber 
Q_{s,k} = \boldsymbol{G}_{s,k:} (f_{s,k}\boldsymbol{e}_{s} - e_{s,k}\boldsymbol{f}_{s}) - \boldsymbol{B}_{s,k:} &(e_{s,k}\boldsymbol{e}_{s} + f_{s,k}\boldsymbol{f}_{s}),\\
&\forall k \in \mathcal{N}_s,
\end{align}
where $\boldsymbol{G}_{s,k:}$ and $\boldsymbol{B}_{s,k:}$ are the the $k^{th}$ row of the conductance and susceptance matrices $\boldsymbol{G_{s}}$ and $\boldsymbol{B_{s}}$. The magnitude of voltage-controlled buses is constrained as follows:
\begin{align}
\label{eqn:AC_V_equal_onstraint}
g_{s,k}^{V}(\boldsymbol{X}) = e_{s,k}^2 + f_{s,k}^2 - {V_{s,k}^{o}}^2 = 0, \hspace{.2cm}\forall k \in {\mathcal{G}_s \cup \mathcal{V}_s}
\end{align}

The inequality constraints in HVac grid $s$ include the generation limits of all generators, which are given as follows:
\begin{align}
\label{eqn:AC_Gen_equal_onstraint}
\nonumber
{\ubar{P}_{s,k}^{gen}} &\le {P}_{s,k}^{gen} \le {\bar{P}_{s,k}^{gen}}, \\
{\ubar{Q}_{s,k}^{gen}} &\le {Q}_{s,k}^{gen} \le {\bar{Q}_{s,k}^{gen}}, \hspace{.2cm}\forall k \in {\mathcal{G}_s}.
\end{align}

The dispatch $Q_{s,k}^{sh}$ of the shunt capacitor at bus $k$ belongs to a set $\mathcal{Q}_{s,k}^{sh}$ with multiple discrete values:
\begin{align}
\label{eqn:Cap_constraints_original}	
Q_{s,k}^{sh} \in \mathcal{Q}_{s,k}^{sh}, \forall k \in \mathcal{N}_s^{sh}.
\end{align}

The voltages at load buses are limited by lower and upper bounds as follows:
\begin{align}
\label{eqn:AC_V_inequal_constraint}
{\ubar{V}}^{2} \le h_{s,k}^{V}(\boldsymbol{X}) = e_{s,k}^2 + f_{s,k}^2 \le {\bar{V}}^{2}, \hspace{.2cm}\forall k \in {\mathcal{L}_s}.
\end{align}
The line flow constraint, which represents the thermal limit of the line between bus $k$ and bus $j$, is given as follows:
\begin{align}
\label{eqn:AC_I_inequal_constraint}
\nonumber
h_{s,kj}^{I}(\boldsymbol{X}) &= (g_{s,kj}^2+b_{s,kj}^2) [{(e_{s,k}-e_{s,j})}^2 + {(f_{s,k}-f_{s,j})}^2] \\
             &\le \bar{I}_{s,kj}^{2}, \hspace{.2cm} \forall (k,j) \in {\mathcal{D}_s}.
\end{align}
To guarantee the system stability, the active power flowing between bus $j$ and bus $k$ is also limited as follows:
\begin{align}
\label{eqn:AC_P_inequal_constraint}
\nonumber
-\bar{P}_{s,kj} \le& h_{s,kj}^{P}(\boldsymbol{X}) = g_{s,kj} (e_{s,k}^2 \!+\! f_{s,k}^2 \!-\! e_{s,k}e_{s,j} \!-\! f_{s,k}f_{s,j}) \\
+ b_{s,kj}&(e_{s,k}f_{s,j}\!-\!f_{s,k}e_{s,j}) \le \bar{P}_{s,kj},  \forall (k,j) \!\in \! {\mathcal{D}_s}.
\end{align}

\subsection{Constraints of LF-HVac Grid $l$}
Assuming LF-HVac grid $l$ does not serve any loads, the equality constraints $\boldsymbol{g}_{l}^{PQ}$ in HVac grid $l$, representing the active and reactive power balance at all buses, are given by:
\begin{align}
\label{eqn:LFAC_PF_eqn}	
\nonumber
g_{l,k}^{P}(\boldsymbol{X}) &= P_{l,k} - (P_{l, k}^{gen} + P_{l,k}^{conv}) = 0, \\
\nonumber
g_{l,k}^{Q}(\boldsymbol{X}) &= Q_{l,k}- (Q_{l, k}^{gen} + Q_{l,k}^{conv}) \\
            & \hspace{1cm}  - ({e_{l,k}}^2+{f_{l,k}}^2)Q_{l,k}^{sh} = 0, \hspace{.2cm}  \forall k \in \mathcal{N}_{l}.
\end{align}
where the real power $P_{l,k}$ and the reactive power $Q_{l,k}$ injected into LF-HVac grid $l$ from bus $k$ in (\ref{eqn:LFAC_PF_eqn}) are obtained with an expression similar to (\ref{eqn:AC_injected_power_eqn}). 

Similar to HVac grid $s$, voltage constraints (\ref{eqn:AC_V_equal_onstraint}) at voltage-controlled buses, wind generation limits (\ref{eqn:AC_Gen_equal_onstraint}),  capacitor dispatch constraint (\ref{eqn:Cap_constraints_original}), line current limits (\ref{eqn:AC_I_inequal_constraint}), and line power limits (\ref{eqn:AC_P_inequal_constraint}) apply for LF-HVac grid $l$.

\subsection{Constraints of BTB Converters}\label{sec:ConverterConstraints}
To simplify the expression, several equations in this section are shown in polar coordinates. With the power and current directions in Fig. \ref{fig:Back-to-Back_Converter_Control}, the currents $\bar{I}_{s,i}$ withdrawn from HVac grid $s$ and $\bar{I}_{l,i}$ injected into LF-HVac grid $l$ as well as the voltages at the shunt capacitors are calculated as follows:
\begin{align}
\label{eqn:_Is_Il_Vfeqn}
\nonumber
\bar{I}_{s, k}\! &=\! {{P_{s,k}^{conv}\!-\!jQ_{s,k}^{conv}} \over {\bar{V}_{s,k}^{*}}}; \bar{V}_{f1, k}\! =\! \bar{V}_{s, k} \!-\! \bar{I}_{s,k}\bar{Z}_{T1,k}, \forall k \in \mathcal{C}_{s},\\
\bar{I}_{l, k}\! &=\! {{P_{l,k}^{conv}\!-\!jQ_{l,k}^{conv}} \over \bar{V}_{l,k}^{*}}; \bar{V}_{f2, k}\! =\! \bar{V}_{l, k} \!+\! \bar{I}_{l,k}\bar{Z}_{T2,k}, \forall k \in \mathcal{C}_{l},
\end{align}
Converter currents and voltages $\bar{I}_{c1,k}$, $\bar{I}_{c2,k}$, $\bar{V}_{c1,k}$, and $\bar{V}_{c2,k}$ at each side of the BTB converter are determined as follows:
\begin{align}
\label{eqn:_Uc_Ic_eqn}
\nonumber
\bar{I}_{c1, k}\! &=\! \bar{I}_{s, k} - \bar{V}_{f1, k}/\bar{Z}_{f1, k}; \hspace{0.1cm} \bar{V}_{c1, k} \!=\! \bar{V}_{f1, k} \!-\! \bar{I}_{c1,k}\bar{Z}_{c1,k}, \forall k \in \mathcal{C}_{s}, \\
\bar{I}_{c2, k}\! &=\! \bar{I}_{l, k} + \bar{V}_{f2, k}/\bar{Z}_{f2, k}; \hspace{0.1cm} \bar{V}_{c2, k} \!=\! \bar{V}_{f2, k} \!+\! \bar{I}_{c2,k}\bar{Z}_{c2,k}, \forall k \in \mathcal{C}_{l}.
\end{align}
The Joule and switching losses $P_{J1,i}$, $P_{sw1,i}$, $P_{J2,i}$, and $P_{sw2,i}$ at each side of the BTB converter are calculated as follows \cite{Jef_1, Jef_2}:
\begin{align}
\label{eqn:Pcon_eqn}
\nonumber
P_{J1,k} &= {I}_{s, k}^{2} R_{T1,k} + {I}_{c1, k}^{2} R_{c1,k}, \forall k \in \mathcal{C}_{s};\\
\nonumber
P_{sw1, k} &= a_0 + a_1I_{c1, k} + a_2I_{c1, k}^{2}, \forall k \in \mathcal{C}_{s};\\
\nonumber
P_{J2,k} &= {I}_{l, k}^{2} R_{T2,k} + {I}_{c2, k}^{2} R_{c2,k}, \forall k \in \mathcal{C}_{l};\\
P_{sw2, k} &= a_0 + a_1I_{c2, k} + a_2I_{c2, k}^{2} , \forall k \in \mathcal{C}_{l},
\end{align}
where $R_{T1,k}$, $R_{T2,k}$, $R_{c1,k}$, and $R_{c2,k}$ are the winding resistances of the transformers and phase reactors while $a_0$, $a_1$, and $a_2$ are given coefficients.

The power balance equation $g_{cl,k}^{P}$ at BTB converter $k$ connecting HVac grid $s$ and LF-HVac grid $l$ is obtained from the relationship between $P_{s,k}^{conv}$, $P_{l,k}^{conv}$, and the Joule and switching losses:
\begin{align}
\label{eqn:mismatch_conv}
g_{c,k}^{P}(\boldsymbol{X})\!=\! P_{l,k}^{conv} \!+\! P_{J2,k} \!+\! P_{sw2,k} \!+\! P_{J1,k} \!+\! P_{sw1,k} \!-\! P_{s,k}^{conv} \!=\! 0.
\end{align}
By substituting (\ref{eqn:_Is_Il_Vfeqn})-(\ref{eqn:Pcon_eqn}) and ignoring the capacitive element (as in the modular multilevel converter technology), (\ref{eqn:mismatch_conv}) can be rewritten in rectangular coordinates as follows:
\begin{align}
\label{eqn:mismatch_conv_final}
\nonumber
&g_{c,k}^{P}(\boldsymbol{X})\!=\! P_{l,k}^{conv} \!+\! 2a_0 \!-\! P_{s,i}^{conv}\\
\nonumber
&+{{{P_{l,k}^{conv}}^2+{Q_{l,k}^{conv}}^2} \over e_{l,k}^2\!+\!f_{l,k}^2} (R_{2,k} \!+\! a_2) \!+\! \sqrt{{{{P_{l,k}^{conv}}^2 \!+\! {Q_{l,k}^{conv}}^2} \over e_{l,k}^2\!+\!f_{l,k}^2}} a_1\\
&+ {{{P_{s,k}^{conv}}^2+{Q_{s,k}^{conv}}^2} \over e_{s,k}^2\!+\!f_{s,k}^2} (R_{1,i} \!+\! a_2) \!+\! \sqrt{{{{P_{s,k}^{conv}}^2 \!+\! {Q_{s,k}^{conv}}^2} \over e_{s,k}^2\!+\!f_{s,k}^2}} a_1  \!=\! 0,
\end{align}
where $R_{1,k}=R_{T1,k}+R_{c1,k}$ and $R_{2,k}=R_{T2,k}+R_{c2,k}$.

For control design and stability analysis, it is important to take into account the capability constraints of each VSC converter. These constraints are normally converted into equivalent constraints of voltage and power at the ac terminal in order to be easily embedded in optimal power flow algorithms.
\par 1) The RMS converter current $I_{c1,k}$ in (\ref{eqn:_Uc_Ic_eqn}) is limited by an upper bound $I_{c,k}^{max}$ to protect switching devices from thermal overheating \cite{Jef_1}:
\begin{align}
\label{eqn:Ic_constraint}
\nonumber
I_{c1,k} &\leq I_{c,k}^{max} \iff   \\
S_{s,k}^{conv} \! \leq  \! \bigg| V_{s,k}^2  \dfrac{1}{\bar{Z}_{f1,k}^* \!+\! \bar{Z}_{T1,k}^*} \!&+\! \bar{V}_{s,k} \bar{I}_{c,k}^{max*} \dfrac{\bar{Z}_{f1,k}}{\bar{Z}_{f1,k}^* \!+\! \bar{Z}_{T1,k}^*}  \bigg|
\end{align}
Without the shunt capacitive element, (\ref{eqn:Ic_constraint}) simplifies to: 
\begin{align}
\label{eqn:Ic_constraint_sim}
\nonumber
S_{s,i}^{conv} &= \sqrt{{(P_{s,k}^{conv})^2+(Q_{s,k}^{conv})^2}} \leq  I_{c,k}^{max} V_{s,k} \iff\\
\nonumber
h_{s,k}^{Iconv}(\boldsymbol{X})&={(P_{s,k}^{conv})}^2  \! + \! {(Q_{s,k}^{conv})}^2 \\
&- {(I_{c,k}^{max})}^2 (e_{s,k}^2\!+\!f_{s,k}^2) \!\leq\! 0, \forall k \! \in \! \mathcal{C}_{s}.
\end{align}
The similar inequality is obtained for $\textup{VSC}_\textup{2}$ as follows:
\begin{align}
\label{eqn:Ic_constraint_sim2}
\nonumber
h_{l,k}^{Iconv}(\boldsymbol{X})&={(P_{l,k}^{conv})}^2 \!+\!{(Q_{l,k}^{conv})}^2 \\
&- {(I_{c,k}^{max})}^2 (e_{l,k}^2 \!+\! f_{l,k}^2) \!\leq\! 0,  \forall k \!\in\! \mathcal{C}_{l}.
\end{align}
\par 2) The rms converter voltage $V_{c1,k}$ is also limited by the voltage $V_{dc,k}$ across the dc-link capacitor to avoid over-modulation \cite{Jef_1,Feng_1}:
\begin{align}
\label{eqn:Vc_constraint_1}
\nonumber
V_{c1,k} \leq k_m & V_{dc,k} \iff h_{s,k}^{Vconv}(\boldsymbol{X}) = \hspace{1cm} \\
\nonumber
[P_{s,k}^{conv} \!&-\! (e_{s,k}^2\!+\!f_{s,k}^2)g_{1,k}]^2 \!+\! [Q_{s,k}^{conv} \!+\! (e_{s,k}^2\!+\!f_{s,k}^2)b_{1,k}]^2 \\
&-\bigg(\dfrac{k_mV_{dc}}{Z_{1,k}}\bigg)^2 (e_{s,k}^2\!+\!f_{s,k}^2) \!\leq\! 0, \forall k \!\in\! \mathcal{C}_{s},
\end{align}
where $(g_{1,k} + jb_{1,k})$ = 1/${\bar{Z}_{1,k}}$ = 1/(${\bar{Z}_{T1,k}}$+${\bar{Z}_{c1,k}}$), while the value of the coefficient $k_m$ depends on the pulse-width modulation (PWM) technique. For sinusoidal and space-vector PWM, which are the two most popular PWM techniques, $k_m$ is equal to 0.61 and 0.71, respectively.

\begin{figure}[t!]
	\centering
	\includegraphics[width = 0.9\columnwidth] {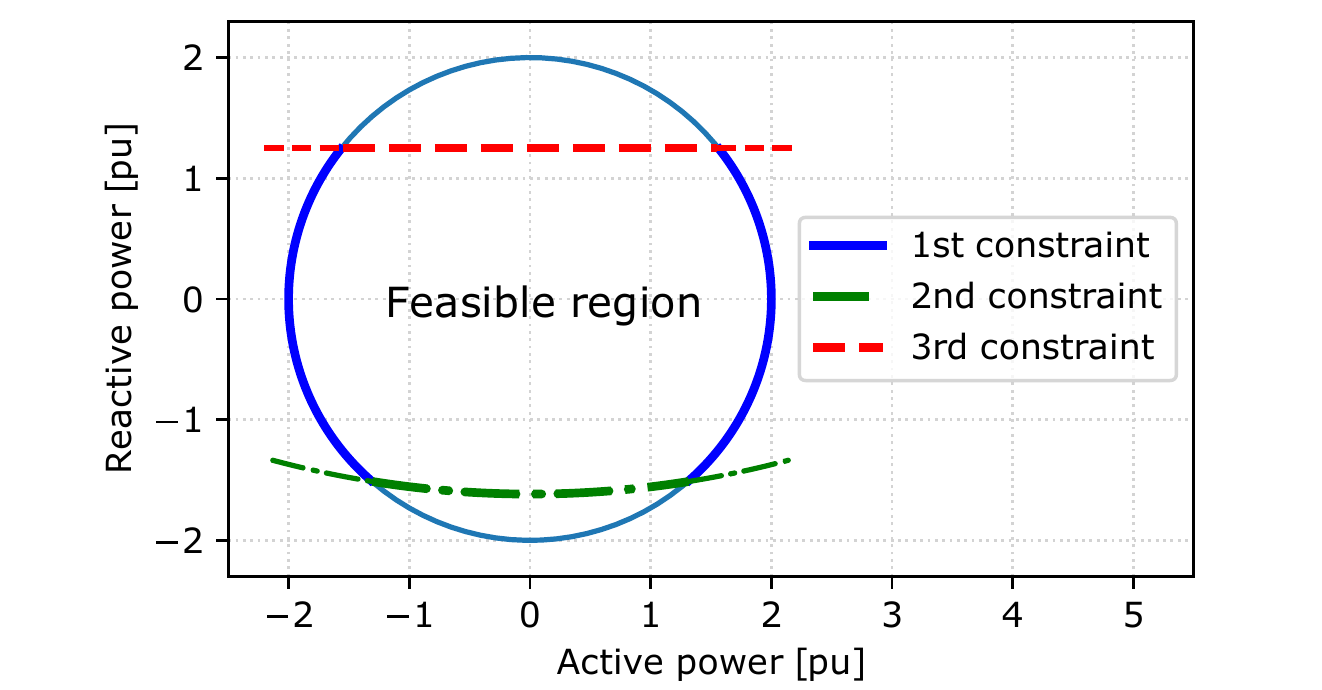}	
	\caption{The feasible operating region of $\textup{VSC}_\textup{1}$ with $V_{s,i}$ = 1pu, $I_{c,i}^{max}$ = 2pu, $k_m$ = 0.61, $\bar{Z}_1$ = 0.0001+j0.15, $S_{rated}$ = 2.5 pu, and $k_Q$ = 0.5.}
	\label{fig:ConverterLimit}
\end{figure}

Noticing the difference in the power convention at two sides of the BTB converter in Fig. \ref{fig:Back-to-Back_Converter_Control}, an inequality is obtained for $\textup{VSC}_\textup{2}$ as follows:
\begin{align}
\label{eqn:Vc_constraint_12}
\nonumber
h_{l,k}^{Vconv}(\boldsymbol{X}) = \hspace{1.5cm} \\
\nonumber
[P_{l,k}^{conv} \!+\! (e_{l,k}^2\!+\!f_{l,k}^2)g_{2,k}]^2 &\!+\! [Q_{l,k}^{conv} \!-\! (e_{l,k}^2\!+\!f_{l,k}^2)b_{2,k}]^2 \\
-\bigg(\dfrac{k_mV_{dc}}{Z_{2,k}}\bigg)^2 & (e_{l,k}^2\!+\!f_{l,k}^2) \!\leq\! 0, \forall k \!\in\! \mathcal{C}_{l}.
\end{align}
\par 3) The reactive power absorbed by the converter is also limited with respect to its rated apparent power $S_{rated}$:
\begin{align}
\label{eqn:Q_constraint}
\nonumber
Q_{s,k}^{conv} &\leq k_Q S_{rated},\\
Q_{l,k}^{conv} &\geq -k_Q S_{rated},
\end{align}
where the  coefficient $k_Q$ is project-specific \cite{ABB_1,Feng_1}.

The convex feasible operating region of converter $\textup{VSC}_\textup{1}$, which is enclosed by the three limits described above, is shown in Fig. \ref{fig:ConverterLimit}.

\subsection{Loss Minimization Problem in Multi-Frequency AC Systems}
Considering the objective and all constraints formulated above, the optimal power flow problem in multi-frequency ac systems implementing BTB converters can be defined as follows:
\begin{subequations}
	\label{eqn:Optimization_form}
	\begin{align}
	\nonumber
	min \hspace{0.3cm} &f(\boldsymbol{x}),  \\
	\label{eqn:Optimization_form_equality}
	s. t. \hspace{0.5cm}\boldsymbol{g(x)} &= \boldsymbol{0}\\  
	\label{eqn:Optimization_form_box_inequality}
	\boldsymbol{x}_{min} &\leq \boldsymbol{\hat{I}x} \leq  \boldsymbol{x}_{max}\\
	\label{eqn:Optimization_form_functional_equality}	
	\boldsymbol{h}_{min} &\leq \boldsymbol{h(x)}  \leq   \boldsymbol{h}_{max}
	\end{align}
\end{subequations}
The equality constraint (\ref{eqn:Optimization_form_equality}) is combined from (\ref{eqn:AC_PF_eqn}), (\ref{eqn:AC_V_equal_onstraint}), (\ref{eqn:LFAC_PF_eqn}), and (\ref{eqn:mismatch_conv_final}). The box inequality constraint (\ref{eqn:Optimization_form_box_inequality}) is combined from (\ref{eqn:AC_Gen_equal_onstraint}), (\ref{eqn:Cap_constraints_original}) , and (\ref{eqn:Q_constraint}). The functional inequality constrain (\ref{eqn:Optimization_form_functional_equality}) is combined from  (\ref{eqn:AC_V_inequal_constraint})-(\ref{eqn:AC_P_inequal_constraint}) and (\ref{eqn:Ic_constraint_sim}) - (\ref{eqn:Vc_constraint_12}). If a lower or upper bound in (\ref{eqn:Optimization_form_box_inequality}) or (\ref{eqn:Optimization_form_functional_equality}) is missing in the original inequality, a sufficiently large dummy constant is inserted. Since the feasible of (\ref{eqn:Optimization_form}) is nonconvex and the variable vector includes discrete variables,  (\ref{eqn:Optimization_form}) is a nonconvex MINLP problem.

\section{Solution Approach}\label{sec:PCPDIPM}
\begin{figure}[b!]
	\centering
	\includegraphics[width = 1\columnwidth] {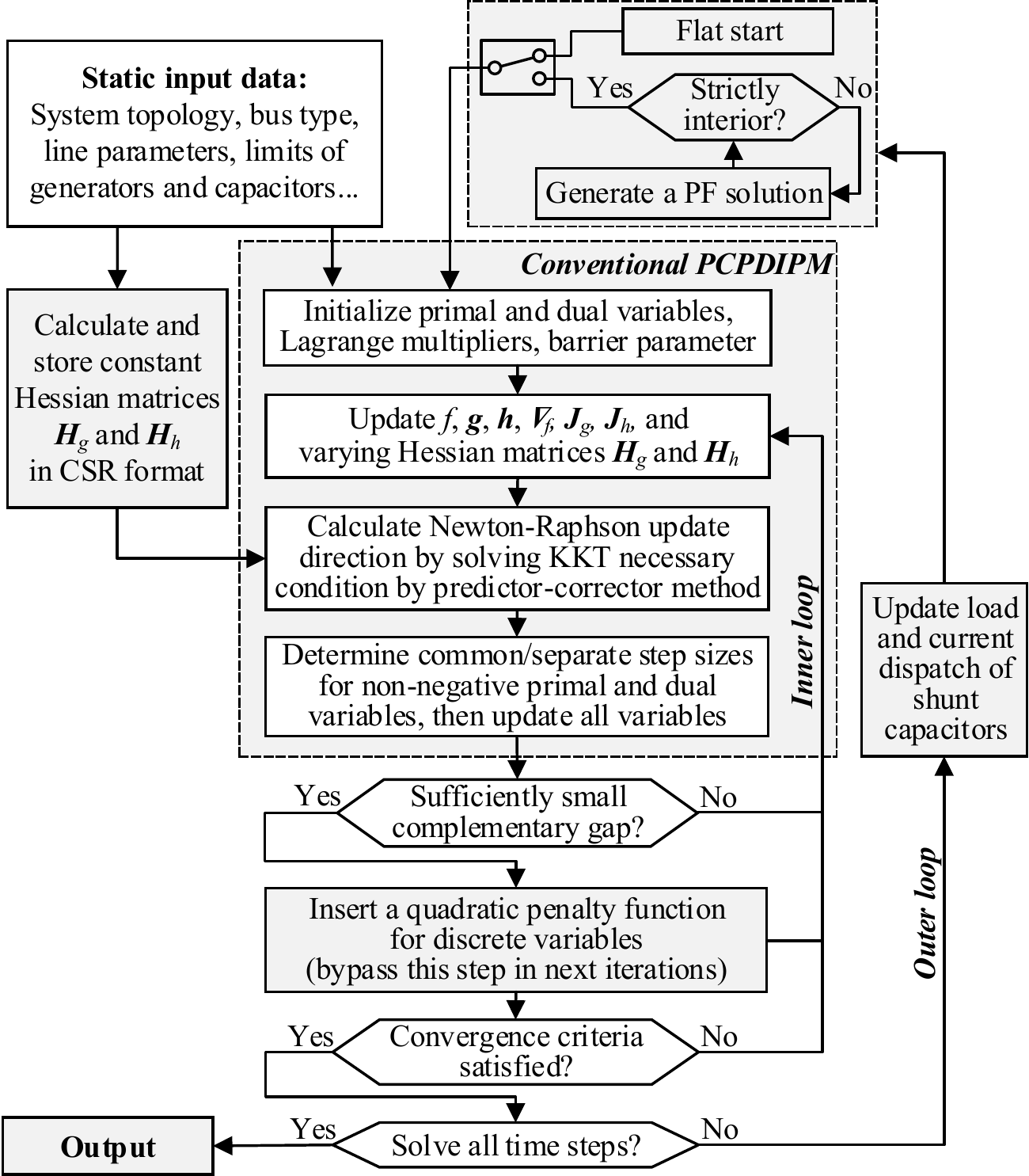}	
	\caption{Flow chart of the modified PCPDIPM method to solve the formulated multi-period OPF problem in a multi-frequency HVac system.}
	\label{fig:Flowchart}
\end{figure}

This section describes an efficient solution approach to solve the formulated multi-period MINLP problem (\ref{eqn:Optimization_form}) in a generalized multi-frequency power system. The flowchart of this modified PCPDIPM method is shown in Fig. \ref{fig:Flowchart}.

The developed framework has two main loops. The inner loop solves for the optimal dispatch of generators, shunt capacitors, and BTB converters in a time step, while the outer loop is the repetition of the inner loop for multiple time steps in a given time horizon. The linkage between two consecutive time steps is that the solution of subsequent time step depends on the shunt capacitor banks in service at the previous time step.

Within a time step, as shown in Fig. \ref{fig:Flowchart}, the solution of (\ref{eqn:Optimization_form}) is obtained based on the PCPDIPM. The PCPDIM is a Newton-Raphson-based method where the variable update in each iteration is determined as the solution of a system of linear equations $\boldsymbol{Ax}=\boldsymbol{b}$. While solving such a system of equations can be done efficiently by using a sparse linear solver, forming $\boldsymbol{A}$ and $\boldsymbol{b}$ in the context of OPF is computationally expensive for a large transmission system. This computational burden results from iteratively updating a high number of the Jacobian and Hessian matrices corresponding to the power balance equations (\ref{eqn:AC_PF_eqn}), (\ref{eqn:LFAC_PF_eqn}), and (\ref{eqn:mismatch_conv_final}). This paper takes the advantages of constant Hessian matrices of all power balance constraints (\ref{eqn:AC_PF_eqn}) and (\ref{eqn:LFAC_PF_eqn}) in HVac and LF-HVac grids by formulating the OPF problem in rectangular coordinates. Therefore, these matrices are pre-calculated once and without any approximations for the entire multi-period OPF. In addition, a compressed sparse row (CSR) format is used for storing only the nonzero entries in these highly sparse and large matrices so that this step is applicable in a normal computer memory capacity. Unlike (\ref{eqn:AC_PF_eqn}) and (\ref{eqn:LFAC_PF_eqn}), both the Jacobian of Hessian matrices of the power balance constraints (\ref{eqn:mismatch_conv_final}) in BTB converter stations are not constant. The exact forms of these matrices are derived and shown in the Appendix, and their values need to be updated every iteration.

Another challenging issue when adopting the conventional PCPDIPM is that it cannot guarantee a global optimum for a nonconvex problem. Therefore multiple starting points, which include both a flat start and warm starts, are used in this work to improve the quality of the final solution. The warm starts, which strictly satisfy all inequality constraints of (\ref{eqn:Optimization_form}), are generated from a separately developed PF tool for multi-frequency power system \cite{Quan_3}. The final solution is chosen as the one with the minimum objective function.

As discussed above, the existence of discrete variables prevents the application of NLP solvers for the OPF problem (\ref{eqn:Optimization_form}). In this work, the technique to deal with discrete variables in \cite{Liu_1} is adopted. At the beginning, all variables are considered to be continuous. When the solution is about to converge, i.e the complementary is sufficiently small, the dispatch of shunt capacitors is forced to converge to the nearest discrete values by adding a penalty function to the objective function. 

\begin{figure}[b]
	\centering
	\includegraphics[width = 1\columnwidth] {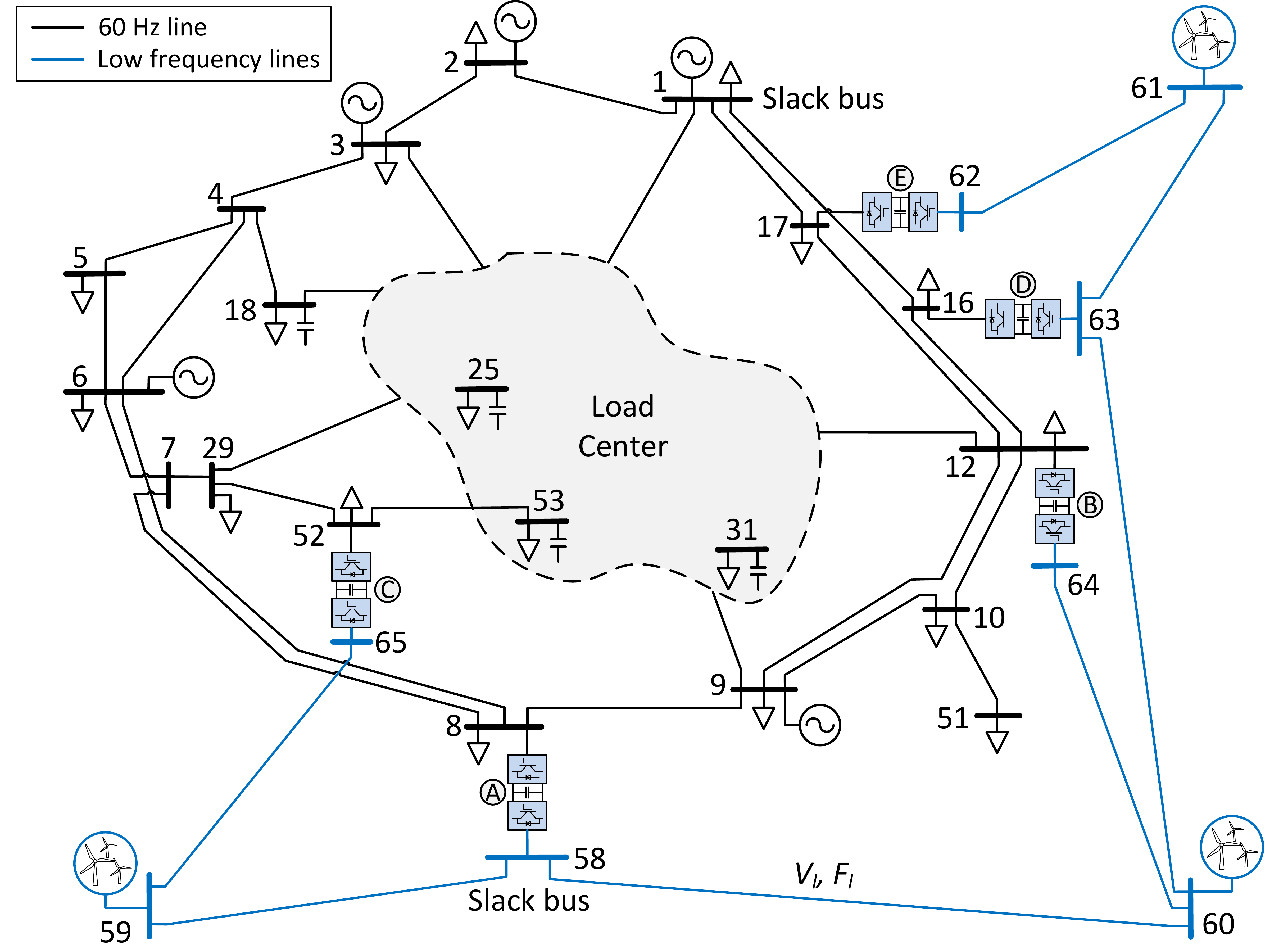}	
	\caption{The multi-frequency power system used to validate the proposed optimal power flow formulation and solution. It consists of 60-Hz HVac and LF-HVac lines interconnected by BTB converters. Adapted from \cite{TestTransmissionSystems}.}
	\label{fig:57bus_System}
\end{figure}
\begin{figure}[t]
	\centering
	\includegraphics[width = 0.8\columnwidth] {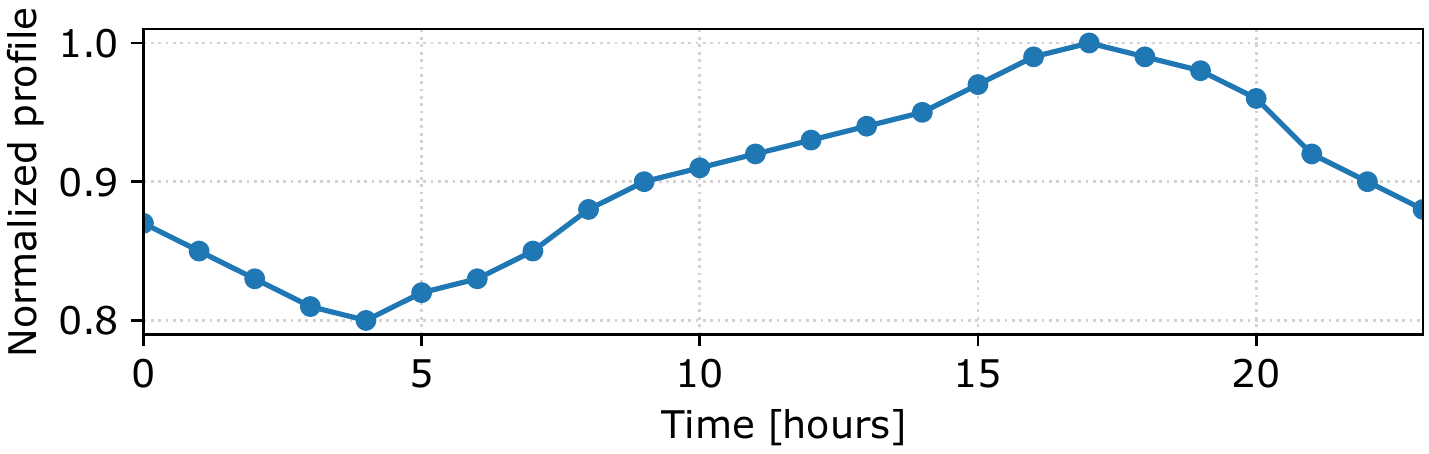}	
	\caption{Normalized 24-hour load profile.}
	\label{fig:LoadProfile}
\end{figure}
\section{Case Study}
This section demonstrates the benefits of the proposed formulation in Section \ref{sec:Formulation} and solution approach in Section \ref{sec:PCPDIPM} in minimizing system losses,  capacitor switching operations, and voltage violations in a multi-frequency HVac system.

\subsection{System Description}
The studied multi-frequency HVac transmission system shown in Fig.~\ref{fig:57bus_System} is modified from the IEEE 57-bus test system \cite{TestTransmissionSystems}. It consists of the original 60-Hz HVac grid and a 10-Hz LF-HVac grid operated at 138 kV and 500 kV, respectively. 

The HVac grid consists of 57 buses, including one slack bus, 4 PV buses, and 37 load buses with a peak demand of 1464.3 MW. The normalized profile of the actual load at all buses are assumed to be similar and shown in Fig. \ref{fig:LoadProfile}. This grid has 78 short and medium transmission lines. Two generators at Bus 8 and 12 in the original grid are removed and replaced by BTB converters A and B. The other BTB converters C, D, and E are connected to PQ Bus 52, 16, and 17, respectively. All of these 5 BTB converters are scheduled to transfer active power from the LF-HVac grid as well as to support reactive power to the HVac grid, as described in Section \ref{sec:Modeling}. Four capacitor banks are located at Bus 18, 25, 31, and 53 as additional reactive power sources with an initial dispatch at their maximum rating of 10.01, 9, 10, and 11.88 Mvar, respectively.

The 10-Hz LF-HVac grid consists of 8 buses and 7 300-km transmission lines. There are 5 onshore BTB converters, and the converter at Bus 58, which is the slack bus of the LF-HVac grid, regulates the voltage at this bus. The converters at offshore wind farms at PV Bus 59-61 are omitted to illustrate an application of direct power generation at low ac frequency. The maximum power generation from offshore wind farms is assumed to be constant in the simulated day.

The switching loss coefficients of the BTB converters described in Section \ref{sec:ConverterConstraints} are assumed to be identical, and they are shown in Table \ref{tab:SwitchingLossCoefficients}. The other operating parameters of the BTB converters, which are used to determine their feasible operating regions, are shown in Table \ref{tab:ConverterParam}.

\begin{table}[t!]
	\setlength{\tabcolsep}{8.5pt}
	\caption{Switching Loss Coefficients of BTB Converters}
	\renewcommand{\arraystretch}{1.1}
	\label{tab:SwitchingLossCoefficients}
	\centering
	\begin{tabular}[h]{|c|ccc|}
		\hline
		Mode  						& $a_0$	& $a_1$	& $a_2$	 \\ 
		\hline								 	 									   
		{Rectifier}        			& 11.033$\times{10^{-3}}$	& 3.464$\times{10^{-3}}$	& 4.400$\times{10^{-3}}$	 \\ 
		{Inverter}        	    	& 11.033$\times{10^{-3}}$	& 3.464$\times{10^{-3}}$	& 6.667$\times{10^{-3}}$	 \\													
		\hline  		
	\end{tabular}
\end{table}
\begin{table}[t!]
	\setlength{\tabcolsep}{4.3pt}	
	\caption{BTB Converter Parameters}
	\renewcommand{\arraystretch}{1.1}
	\label{tab:ConverterParam}
	\centering
	\begin{tabular}[h]{|c|ccccccc|}
		\hline
		& $V_{dc}$ & {$K_{dc}$} 	& {$R_{1},R_{1}$}	& {$X_{1},X_{2}$}	& {$I_{cmax}$} &	{$S_{rated}$} &  $K_{Q}$\\ 
		Converter   				& [kV]	& 	& {[pu]}	& {[pu]}	& {[pu]} &	{[MVA]} & \\ 
		\hline								 	 									   
		$A,B$       	    & 70	& 0.61	& 0.0001	& 0.08	&  3 & 300 & 0.5		 \\ 
		$C-E$    	    & 60    & 0.71  & 0.0001	& 0.08  &  2  & 200 & 0.5		 \\													
		\hline  		
	\end{tabular}
\end{table}

\subsection{Numerical Results}
The system performances in the following cases are discussed and compared:
\begin{itemize}[leftmargin=*]
	\item Case 1: OPF is disabled, and the system operates based on a given power dispatch of generators, shunt capacitors, and BTB converters. With the variation of the load in the simulated day, the operating points of the system are determined from a PF solver developed for multi-frequency system \cite{Quan_3}.
	\item Case 2: OPF is enabled with only generators and shunt capacitors as control resources. To relax the problem, capacitor switching penalty is not included in the objective function by choosing the weighted coefficients ($\alpha_1, \alpha_2$) = (1, 0.0).
	\item Case 3: OPF is enabled with all available power dispatch resources, i.e. generators, shunt capacitors, and BTB converters. To demonstrate the superiority of converter optimal control, capacitor switching is penalized by choosing ($\alpha_1, \alpha_2$) = (1, 0.2).
\end{itemize}

\begin{figure}[t!]
	\centering	
	\includegraphics[width = 0.9\columnwidth] {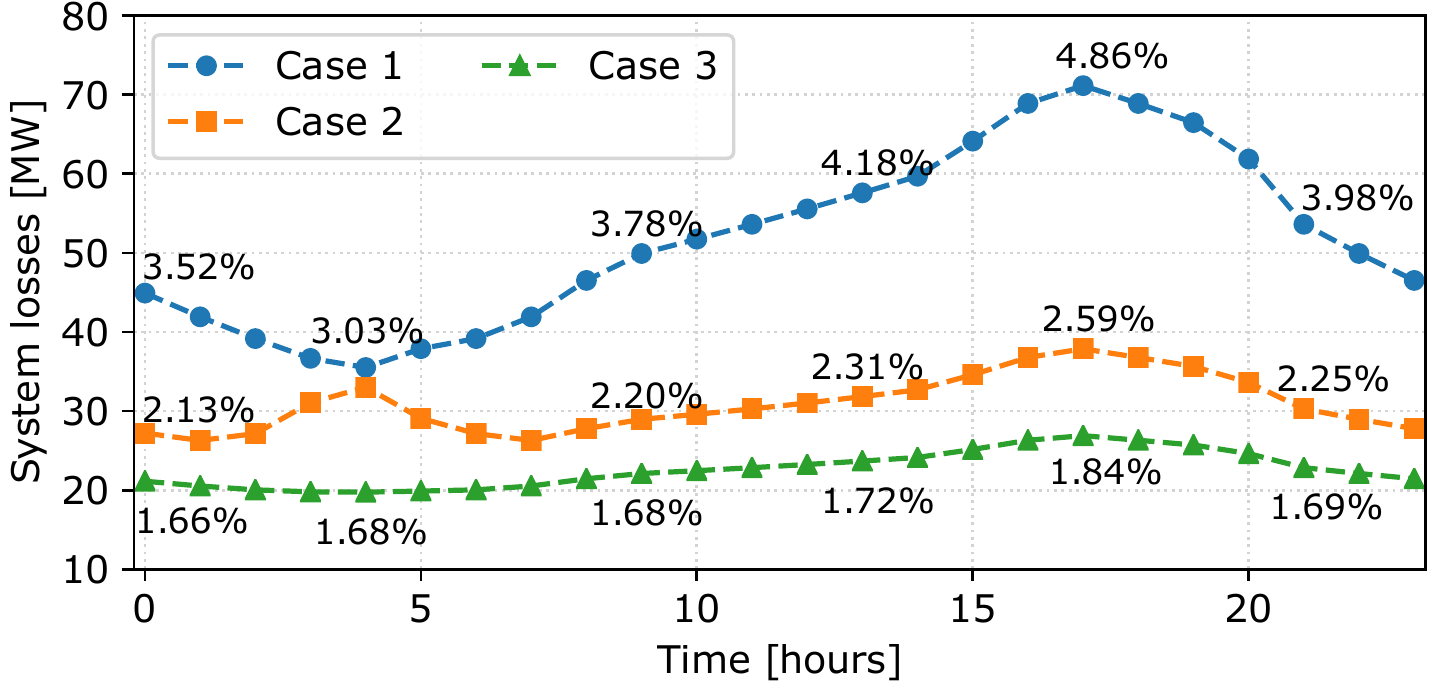}		
	\caption{System losses in the three studied cases in the simulated day.}	
	\label{fig:Losses}
\end{figure}

Fig. \ref{fig:Losses} shows the MW losses of the multi-frequency power system in three cases and the corresponding percentages compared to the demand. It can be seen that the system losses are highest when OPF is disabled in Case 1 and lowest in Case 3 with all optimal power control resources. At the peak load, the system losses reduce from 4.86\% in Case 1 to 2.59\% and 1.84\% in Cases 2 and 3, respectively. In addition, system losses in Case 3 are less sensitive to load changes with the optimal dispatch.

%\textcolor{red}{Next version: Add one bar figure that breaks down HVac transmission loss, LF-HVac loss, converter loss.}

\begin{figure}[t!]
	\centering	
	\includegraphics[width = 0.85\columnwidth] {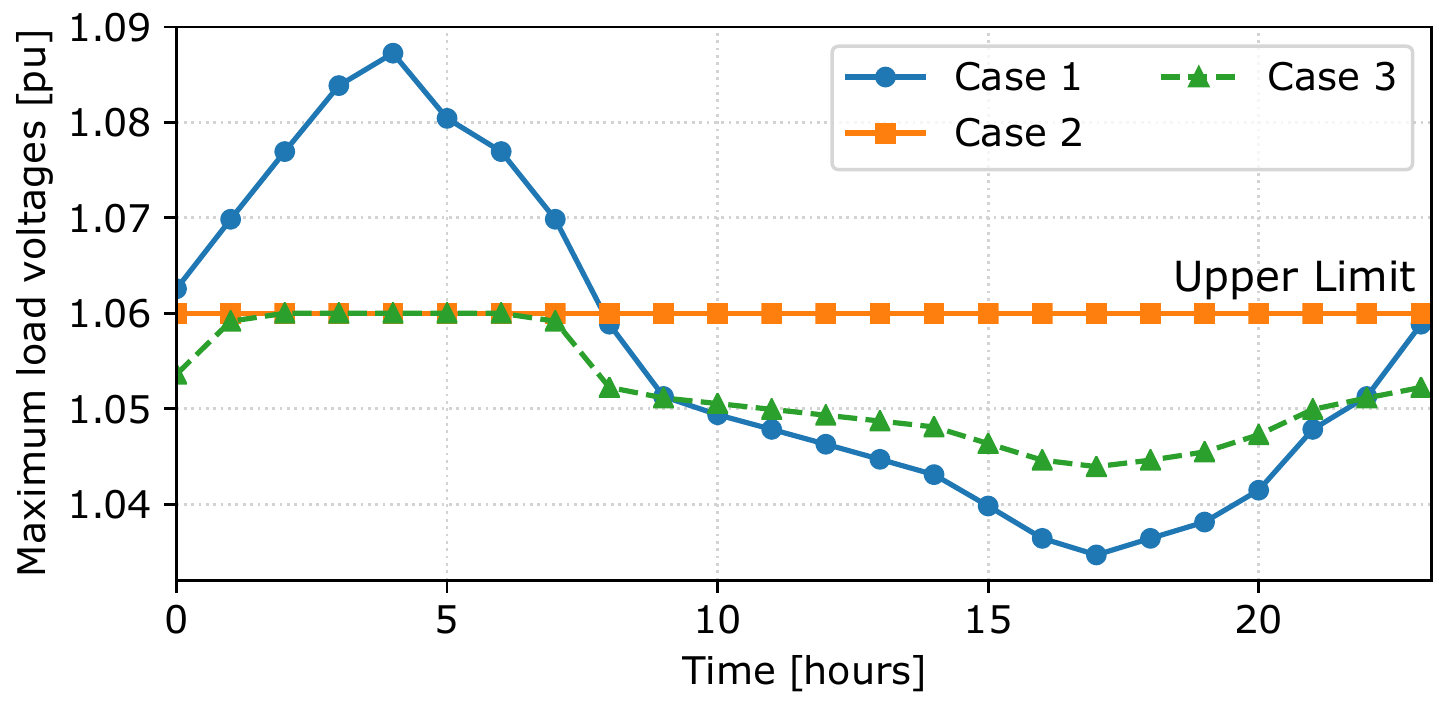}	
	\caption{Maximum load voltages in the three cases in the simulated day.}	
	\label{fig:Vload_max}
\end{figure}

Fig. \ref{fig:Vload_max} shows the maximum voltages at all load buses in the studied cases during the simulated day. While overvoltage appears in Case 1 with a chosen upper limit of 1.06 pu, the optimal dispatch of the generators, shunt capacitors, and BTB converters eliminates voltage violation in the system through out the day in both Cases 2 and 3.

\begin{figure}[t!]
	\centering	
	\includegraphics[width = 0.83\columnwidth] {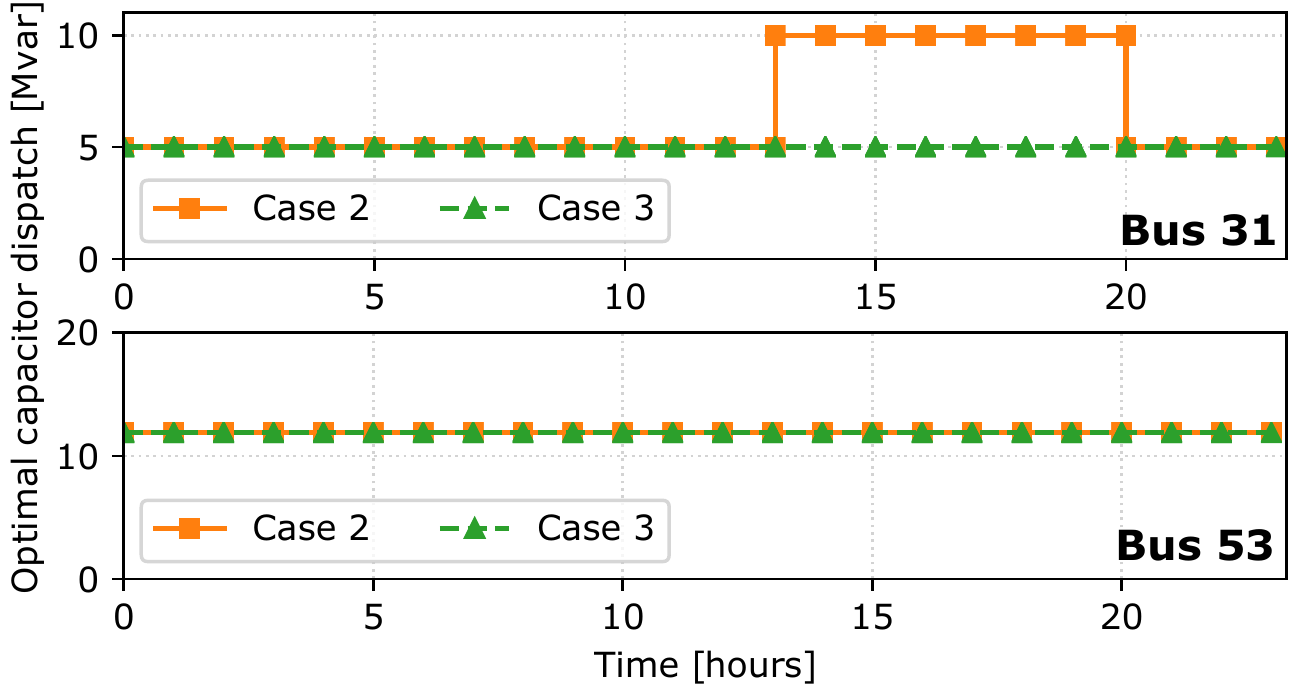}		
	\caption{The optimal dispatch of the shunt capacitors at Bus 31 and 53 in Case 2 (without capacitor switching penalty) and Case 3 (with a capacitor switching penalty).}	
	\label{fig:CapacitorDispatch}
\end{figure}
Fig. \ref{fig:CapacitorDispatch} shows the dispatch of the shunt capacitors at Bus 31 and 53 in Cases 2 and 3 without and with a penalty for capacitor switching, respectively. Besides lower system losses as shown in Fig. \ref{fig:Losses}, less capacitor switching operations at Bus 31 are needed in Case 3 compared to Case 2. The dispatch of the capacitors at Bus 18, 25, and 53 are constant at their respective maximum values. In addition, the dispatch of all capacitors converges exactly at their available discrete values at all time steps, which validates the efficiency of the solution approach in Section \ref{sec:PCPDIPM} in solving the formulated MINLP OPF.

\begin{figure}[t!]
	\centering	
	\includegraphics[width = 0.85\columnwidth] {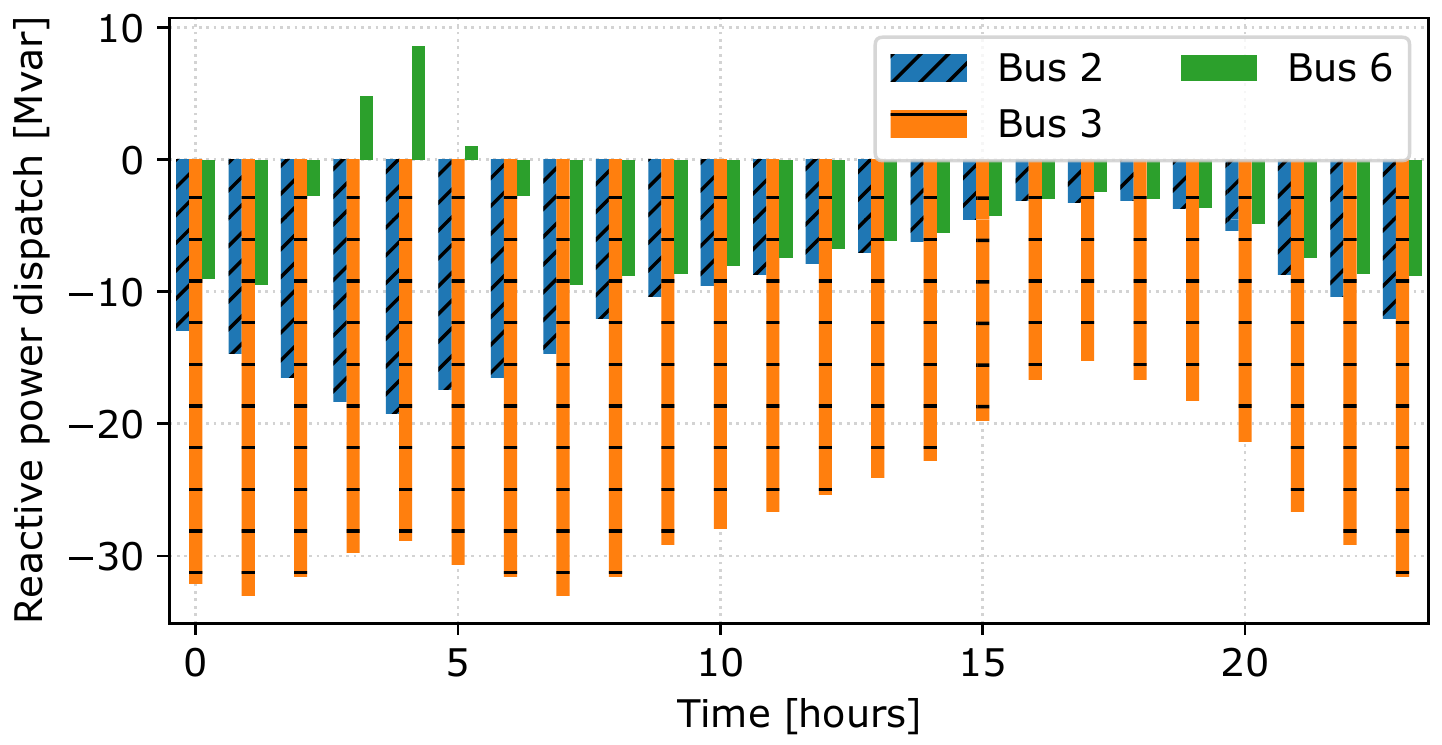}		
	\caption{Optimal reactive dispatch of the generators at Bus 3, 6, and 9 in Case 3.}
	\label{fig:Qsgen}
\end{figure}

Fig. \ref{fig:Qsgen} shows the optimal dispatch of the generators at Bus 2, 3, and 6 in Case 3 during the simulated day. These profiles are consistent with the load characteristic shown in Fig. \ref{fig:LoadProfile}. During the light load periods, the generators absorb reactive power to avoid overvoltages in the HVac grid. In the remaining periods when the demand is higher, the generators reduce reactive power absorption or inject reactive power to the HVac grid to increase grid voltage.

\iffalse
\begin{table}[t!]     
	\setlength{\tabcolsep}{4pt}	                                                   
	\caption{Optimal Power Dispatch of BTB Converters in Case 3 at 5 pm.}
	\renewcommand{\arraystretch}{1.2}
	\label{tab:ConverterDP_3}
	\centering
	\begin{tabular}[h]{|c|ccc|ccc|c|}
		\hline
			& \multicolumn{3}{|c|}{From VSC$_1$ to HVac}  	& \multicolumn{3}{|c|}{From LF-HVac to VSC$_2$ } & \\ \cline{2-7} 
		Converters  & {$P_s^{conv}$}   	& {$Q_s^{conv}$} & {$V_s$}	  	& {$P_l^{conv}$}	& {$Q_l^{conv}$} 	& {$V_l$}  &  Loss\\
		& {(MW)}   	& {(Mvar)} 	&{(pu)}  	& {(MW)}	& {(Mvar)} 	&{(pu)}  & {(\%)}   \\		
		\hline								 	 									   
		$A$         & 177.58    		& 23.58    	&	0.9827 	&  \textcolor{red}{178.71}		&  \textcolor{red}{-248.39}  &   1.02  & 0.63	\\ 
		$B$         & \textcolor{red}{296.30}       & \textcolor{red}{39.82}    & 0.9965	&  297.89		&  14.62   &	1.0059 & 0.53	\\ 
		$C$         & 46.96  		& 5.87    	&	1.0139	&  47.25		&  32.82 &  1.0088	& 0.61 \\ 
		$D$         & \textcolor{red}{199.79}       & \textcolor{red}{44.46}    &	1.0234	& \textcolor{red}{200.67} 			&  \textcolor{red}{31.79}    & 	1.0096	& 0.44\\ 
		$E$         & 167.91  		& 46.13     &	1.0341	& 168.62  			&  18.90    &  1.0079 &	0.42\\ 	
		\hline  		
	\end{tabular}
\end{table}
\fi

\begin{table}[t!]     
	\setlength{\tabcolsep}{4pt}	                                                 
	\caption{Optimal Power Dispatch of BTB Converters in Case 3 at 5 pm.}
	\renewcommand{\arraystretch}{1.15}
	\label{tab:ConverterDP_3}
	\centering
	\begin{tabular}[h]{|c|ccc|ccc|}
		\hline
		& \multicolumn{3}{|c|}{From HVac to VSC$_1$}  	& \multicolumn{3}{|c|}{From LF-HVac to VSC$_2$} \\ \cline{2-7} 
		Converters  & {$P_s^{conv}$}   	& {$Q_s^{conv}$} & {$V_s$}	  	& {$P_l^{conv}$}	& {$Q_l^{conv}$} 	& {$V_l$} \\
		& {(MW)}   	& {(Mvar)} 	&{(pu)}  	& {(MW)}	& {(Mvar)} 	&{(pu)}    \\		
		\hline								 	 									   
		$A$         & -177.58    		& -25.07    	& 0.9829	 	&  \textcolor{red}{178.71}		&  \textcolor{red}{-248.39}  &   1.02	\\ 
		$B$         & \textcolor{red}{-296.14}       & \textcolor{red}{-41.19}    & 0.9966	&  297.73		&  14.63   &	1.0059 	\\ 
		$C$         & -46.96  		& -6.19    	&	1.0395	&  47.24		&  32.82 &  1.0088 \\ 
		$D$         & \textcolor{red}{-199.74}       & \textcolor{red}{-45.03}    &	1.0237	& \textcolor{red}{200.62} 			&  \textcolor{red}{22.86}    & 	1.0096	\\ 
		$E$         & -168.11  		& -46.90     &	1.0347	& 168.81  			&  18.90    &  1.0079\\ 	
		\hline  		
	\end{tabular}
\end{table}
\begin{figure}[t!]
	\centering	
	\includegraphics[width = 1\columnwidth] {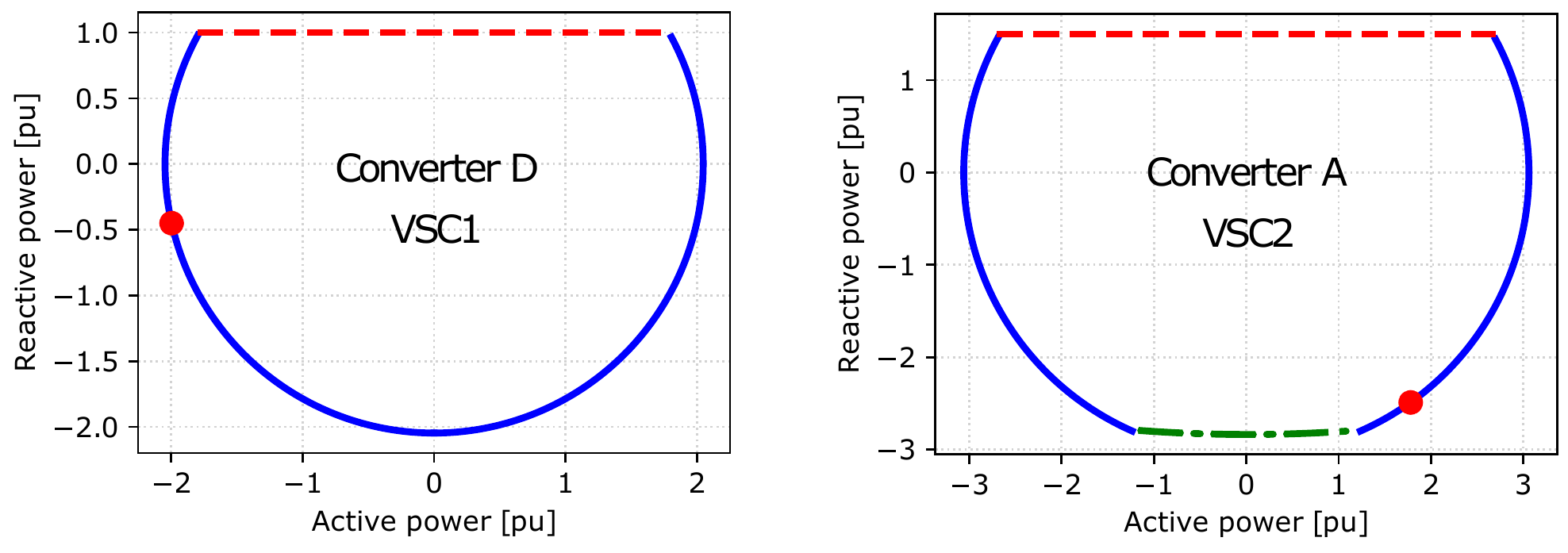}		
	\caption{Feasible operating regions and optimal operating points at the $\textup{VSC}_1$ and $\textup{VSC}_2$ sides of BTB converters $D$ and $A$, respectively, at 5 pm.}
	\label{fig:VSCLimit}
\end{figure}
Table \ref{tab:ConverterDP_3} shows the optimal dispatch at both sides of the BTB converters and the corresponding ac voltages at the points of connection to the HVac and LF-HVac grids during the peak load in Case 3. Converter losses can also be obtained from this table. The red numbers denote the binding to the constraints described in \ref{sec:ConverterConstraints} of converter dispatch. The feasible operating regions and the binding operating points at the $\textup{VSC}_1$ and $\textup{VSC}_2$ sides of converters $D$ and $A$ are shown in Fig. \ref{fig:VSCLimit}.

\subsection{Convergence Analysis}
The proposed solution approach in Section \ref{sec:PCPDIPM} is implemented in Python on an Intel Core i7-6700 processor with 32GB RAM to solve the OPF problem (\ref{eqn:Optimization_form}). The convergence tolerance is set to $10^{-6}$ pu for solution feasibility, objective function, and complementary gap. A flat start is used for both solving PF in Case 1 and OPF in Cases 2 and 3.

As shown in Fig. \ref{fig:Flowchart}, the constant Hessian matrices corresponding to the power balance constraints in rectangular coordinates are pre-computed and retrieved in each iteration. For the system shown in Fig. \ref{fig:57bus_System}, such an approach reduces approximately 90\% of the computational time in each iteration in the modified PCPDIPM. Besides conventional constraints described in Section \ref{sec:Formulation}, the exact Jacobian and Hessian matrices of converter constraints (\ref{eqn:mismatch_conv_final}) - (\ref{eqn:Q_constraint}), derived and shown in the Appendix, can be incorporated in a solver to improve its efficiency and robustness \cite{IPOPT,KNITRO}.

The computational times for 24 time steps and the corresponding average times per step in the three studied cases are shown in Table \ref{tab:CPUTime}. The time in Case 1 is lowest since it is just for solving power flow. In all instances in Cases 2 and 3, the PCPDIPM performs well. The average run time per step of the most generalized OPF problem in Case 3 is approximately 1 second, which allows the proposed formulation and solution approach to be applied to larger power systems.

\setlength{\tabcolsep}{4pt}
\begin{table}[t!]
	\renewcommand{\arraystretch}{1.3}	;
	\caption{Computational times in three studied cases using the proposed method.}
	\label{tab:CPUTime}
	\centering
	\begin{tabular}[h]{|c|cc|cc|cc|}
	\hline
	Case & \multicolumn{2}{|c|}{Case 1 (PF)} & \multicolumn{2}{|c|}{Case 2 (OPF)} & \multicolumn{2}{c|}{Case 3 (OPF)} \\  \cline{1-7}
	Number of steps		& {24 steps}   & {1 step}  & {24 steps}  & {1 step} & {24 steps} & {1 step}  \\
	\cline{1-7}	 		
	Running time (s)	& 0.66	& 0.027  & 18.20	& 0.76 & 27.36	& 1.14  	  \\				
	\hline  		
\end{tabular}
\end{table}
\setlength{\tabcolsep}{3.7pt}
\begin{table}[t!]
	\renewcommand{\arraystretch}{1.3}	
	\caption{Comparison of objective function value and number of iterations using the proposed modified PCPDIPM method and open-access solvers.}
	\label{tab:Solver_Comparison}
	\centering
	\begin{tabular}[h]{|cc|cc|cc|cc|}
		\hline
		\multicolumn{2}{|c|}{}				& \multicolumn{2}{|c|}{IPOPT}  					& \multicolumn{2}{|c|}{BONMIN} 					& \multicolumn{2}{c|}{Modified PCPDIPM} 	\\
		\multicolumn{2}{|c|}{System}		& \multicolumn{2}{|c|}{(NLP)}  					& \multicolumn{2}{|c|}{(MINLP)} 				& \multicolumn{2}{c|}{(MINLP)} 				\\		  \cline{1-8}
		60-Hz		& 	LF-					& 	Objective  	&  Iters    		& Objective   	&  Iters 		& Objective    		&  Iters 					\\
		HVac		& 	HVac 				& (MW)   	 	&    		& (MW)   		& 				& (MW)   		&   	\\
		\cline{1-8}	 			
		3-bus 		& 2-bus					& 70.42			& 	40		& 70.42			& 	63			& 70.42			& 11		\\	
		57-bus		& 8-bus					& 1,505.84		&  	99		& 1,505.97		& 	138			& 1,505.17		&	24  \\							
		\hline  		
	\end{tabular}
\end{table}

Table \ref{tab:Solver_Comparison} shows a comparison about the objective function and number of iterations of the modified PCPDIPM method, NLP solver IPOPT, and MINLP solver BONMIN \cite{IPOPT,BONMIN}. The comparison is applied to two multi-frequency systems, including the system shown in Fig. \ref{fig:57bus_System}, without capacitor switching penalty. It can be seen that IPOPT converges faster than BONMIN, but the capacitor dispatch is unable to converge to their practical discrete values. Although the capacitor dispatch converges exactly to discrete values when using BONMIN, this solver requires the highest number of iterations to converge in both studied systems. The modified PCPDIPM described in Section \ref{sec:PCPDIPM} requires the lowest number of iterations and also guarantees the discrete convergence of capacitor dispatch as BONMIN. Such an improvement is due to the use of exact Hessian matrices and predictor-corrector method. The solutions, including objective function and all continuous and discrete variables, from the modified PCPDIPM and BONMIN closely match each other.

\iffalse
\setlength{\tabcolsep}{3.7pt}
\begin{table}[t!]
	\renewcommand{\arraystretch}{1.3}	
	\caption{Comparison of objective function values and run times using the modified PCPDIPM method and open-access solvers.}
	\label{tab:Solver_Comparison}
	\centering
	\begin{tabular}[h]{|c|c|c|c|}
		\hline
		\multicolumn{2}{|c|}{Studied systems}			& {3-bus HVac and}  					& {57-bus HVac and }	\\	
		\multicolumn{2}{|c|}{}				& {2-bus LF-HVac}  					& {8-bus HVac}	\\	  \cline{1-4}
		\cline{1-4}	 			
		IPOPT 		& Objective (MW)		& 70.42			& 	1,505.84		\\	
		(NLP)		& Iterations			& 40			&  	99		\\							
					& Time (s)				& 0.14			&  	0.46	  	\\		
		\cline{1-4}	 			
		BONMIN 		& Objective (MW)		& 70.42			& 	1,505.97	\\	
		(MINLP)		& Iterations			& 63			& 138			\\							
					& Time (s)				& 0.44			&  	1.06	  	\\	
		\cline{1-4}	 			
		Modified 	& Objective (MW)		& 70.42			& 	1,505.17	\\	
		PCPDIPM		& Iterations			& 11			&  	24		\\							
		(MINLP)		& Time (s)				& 0.19			&  	1.75	  	\\								
		\hline  		
	\end{tabular}
\end{table}  
\fi

%\textcolor{red}{Next version: Add one plot about gap vs. iterations.}
\section{Conclusion}
This paper proposes a multi-period OPF formulation and solution for a multi-frequency HVac transmission system that interconnects conventional 50/60-Hz and low-frequency grids using BTB converters. Besides transferring bulk power from remote generation units, the converters are considered as control devices in addition to the existing generators and shunt capacitors to minimize system losses, capacitor switching operations, and voltage violations. A comprehensive MINLP formulation is proposed, taking into account the control and losses of BTB converters. The formulated problem is efficiently solved by a developed framework based on the PCPDIPM algorithm, addressing the discrete nature of variables. The proposed approach is verified by a multi-frequency system modified from the IEEE 57-bus system. With the obtained optimal dispatch, the results show a significant 3\% loss reduction during peak load condition and elimination of voltage violations through out the simulated day. The solution framework performs fast and robustly in all studied cases and time steps due to the use of the exact forms of the Jacobian and Hessian matrices corresponding to the operational constraints.

\section{Appendix}
This section shows the exact symmetric Hessian matrices corresponding to constraints (\ref{eqn:mismatch_conv_final}) - (\ref{eqn:Q_constraint}), while the corresponding Jacobian matrices can be found from the PF study in \cite{Quan_3}. Since the formulations are similar for both $\textup{VSC}_\textup{1}$ and $\textup{VSC}_\textup{2}$ sides, the grid notations $s$ and $l$ associated with 60-Hz HVac and LF-HVac grids, respectively, are dropped. These formulations can also be adopted directly in OPF analysis of hybrid HVac - HVdc systems implementing single-stage ac/dc VSC converters.
	\begin{itemize}[leftmargin=*]
		\item Constraint (\ref{eqn:mismatch_conv_final}):
	\end{itemize} 		
		\begin{align}
			\nonumber
			\frac{\partial^2 g_{c,k}^P}{\partial^2 e_{k}} &= \bigg( \dfrac{8e_k^2}{V_k^6} - \dfrac{2}{V_k^4}  \bigg) |S_k^{conv}|(R_{k} + a_2)\\
			\nonumber
			 											& \hspace{0.2cm}	+ a_1\bigg( \dfrac{3e_k^2}{V_k^5} - \dfrac{1}{V_k^3}  \bigg) {|S_k^{conv}|}^2,\\
			\nonumber
			\frac{\partial^2 g_{c,k}^P}{\partial^2 f_{k}} &= \bigg( \dfrac{8f_k^2}{V_k^6} - \dfrac{2}{V_k^4}  \bigg) |S_k^{conv}|(R_{k} + a_2)\\
			\nonumber
														& \hspace{0.2cm} + a_1\bigg( \dfrac{3f_k^2}{V_k^5} - \dfrac{1}{V_k^3}  \bigg) {|S_k^{conv}|}^2,	\\
			\nonumber
			\frac{\partial^2 g_{c,k}^P}{\partial e_{k} \partial f_{k}} &= \dfrac{8e_kf_k}{V_k^6}|S_k^{conv}|(R_{1,k} \!+\! a_2) \!+\! a_1 \dfrac{3e_kf_k}{V_k^5} {|S_k^{conv}|}^2,\\
			\nonumber
			\frac{\partial^2 g_{c,k}^P}{\partial e_{k} \partial P_k^{conv}} &=  -\dfrac{4e_k}{V_k^4}P_k^{conv}(R_{1,k} + a_2) - a_1 \dfrac{e_k}{V_k^3} \dfrac{P_k^{conv}}{{|S_k^{conv}|}^2	},\\	
			\nonumber
			\frac{\partial^2 g_{c,k}^P}{\partial e_{k} \partial Q_k^{conv}} &= -\dfrac{4e_k}{V_k^4}Q_k^{conv}(R_{1,k} + a_2) - a_1 \dfrac{e_k}{V_k^3} \dfrac{Q_k^{conv}}{{|S_k^{conv}|}^2	},	\\		
			\nonumber
			\frac{\partial^2 g_{c,k}^P}{\partial f_{k} \partial P_k^{conv}} &= -\dfrac{4f_k}{V_k^4}P_k^{conv}(R_{1,k} + a_2) - a_1 \dfrac{f_k}{V_k^3} \dfrac{P_k^{conv}}{{|S_k^{conv}|}^2	},	\\
			\nonumber
			\frac{\partial^2 g_{c,k}^P}{\partial f_{k} \partial Q_k^{conv}} &= -\dfrac{4f_k}{V_k^4}Q_k^{conv}(R_{1,k} + a_2) - a_1 \dfrac{f_k}{V_k^3} \dfrac{Q_k^{conv}}{{|S_k^{conv}|}^2	},																				
		\end{align} 
		\begin{align}			
			\nonumber
			\frac{\partial^2 g_{c,k}^P}{\partial^2 P_k^{conv}} &= \dfrac{2}{V_k^2}(R_{1,k} + a_2) + a_1 \dfrac{{Q_k^{conv}}^2}{V_k {|S_k^{conv}|}^6	},\\
			\nonumber
			\frac{\partial^2 g_{c,k}^P}{\partial^2 Q_k^{conv}} &= \dfrac{2}{V_k^2}(R_{1,k} + a_2) + a_1 \dfrac{{P_k^{conv}}^2}{V_k {|S_k^{conv}|}^6	},	\\
			\nonumber
			\frac{\partial^2 g_{c,k}^P}{\partial P_k^{conv} \partial Q_k^{conv}} &= \frac{\partial^2 g_{c,k}^P}{\partial Q_k^{conv} \partial P_k^{conv}}
			\nonumber
			= - a_1 \dfrac{P_k^{conv}Q_k^{conv}}{V_k {|S_k^{conv}|}^6	}	,	
		\end{align} 
where:
	\begin{align}
		\nonumber
		V_k &= \sqrt{e_k^2+ f_k^2},\hspace{0.2cm}
		|S_k^{conv}|=\sqrt{{(P_k^{conv})^2+(Q_k^{conv})^2}}.			
	\end{align} 
	\begin{itemize}[leftmargin=*]
		\item Constraints (\ref{eqn:Ic_constraint_sim}) and (\ref{eqn:Ic_constraint_sim2}):
	\end{itemize} 	
	\begin{align}
		\nonumber
		\frac{\partial^2 h_{k}^{Iconv}}{\partial^2 P_k^{conv}} &= \frac{\partial^2 h_{k}^{Iconv}}{\partial^2 Q_k^{conv}} = 2,\\
		\nonumber	
		\frac{\partial^2 h_{k}^{Iconv}}{\partial^2 e_k}	 &= \frac{\partial^2 h_{k}^{Iconv}}{\partial^2 f_k} = 2 {(I_{c,k}^{max})}^2.
	\end{align}

	\begin{itemize}[leftmargin=*]
		\item Constraints (\ref{eqn:Vc_constraint_1}) and (\ref{eqn:Vc_constraint_12}):
	\end{itemize} 	
	\begin{align}
	\nonumber
	\frac{\partial^2 h_{k}^{Vconv}}{\partial^2 P_k^{conv}} &= \frac{\partial^2 h_{k}^{Vconv}}{\partial^2 Q_k^{conv}} = 2,\\
	\nonumber	
	\frac{\partial^2 h_{k}^{Vconv}}{\partial^2 e_k}	 &=  -4g_{k} \bigg[ P_k^{conv} - g_k(3e_k^2+f_k^2)   \bigg]\\
	\nonumber
						 &\hspace{0.4cm}+ 4b_{k} \bigg[ Q_k^{conv} + b_k(e_k^2+3f_k^2) - 2 k_V^2   \bigg],\\
	\nonumber	
	\frac{\partial^2 h_{k}^{Vconv}}{\partial^2 f_k}	 &=  -4g_{k} \bigg[ P_k^{conv} - g_k(e_k^2+3f_k^2)   \bigg]\\
	\nonumber
						&\hspace{0.4cm}+ 4b_{k} \bigg[ Q_k^{conv} + b_k(3e_k^2+f_k^2) - 2 k_V^2   \bigg],	\\
	\nonumber
	\frac{\partial^2 h_{k}^{Vconv}}{\partial e_{k} \partial f_{k}} &= \frac{\partial^2 h_{k}^{Vconv}}{\partial f_{k} \partial e_{k}} = 8(g_k^2+b_k^2)e_kf_k,	
	\end{align}
	\begin{align}	
	\nonumber
	\frac{\partial^2 h_{k}^{Vconv}}{\partial e_{k} \partial P_k^{conv}}	&= -4g_ke_k, \hspace{0.2cm}
	\frac{\partial^2 h_{k}^{Vconv}}{\partial f_{k} \partial P_k^{conv} } = -4g_kf_k,\\	
	\nonumber
	\frac{\partial^2 h_{k}^{Vconv}}{\partial e_{k} \partial Q_k^{conv}}	&= -4g_kf_k, \hspace{0.2cm}
	\frac{\partial^2 h_{k}^{Vconv}}{\partial f_{k} \partial Q_k^{conv} } = 4b_kf_k,
	\end{align}

where: $k_V = \dfrac{k_mV_{dc}}{Z_{k}}$.

	\begin{itemize}[leftmargin=*]
		\item Constraints (\ref{eqn:Q_constraint}): All Hessian matrices are zero matrices.
	\end{itemize}

\bibliography{QuanNguyen_References}

% Generated by IEEEtran.bst, version: 1.14 (2015/08/26)
\begin{thebibliography}{10}
\providecommand{\url}[1]{#1}
\csname url@samestyle\endcsname
\providecommand{\newblock}{\relax}
\providecommand{\bibinfo}[2]{#2}
\providecommand{\BIBentrySTDinterwordspacing}{\spaceskip=0pt\relax}
\providecommand{\BIBentryALTinterwordstretchfactor}{4}
\providecommand{\BIBentryALTinterwordspacing}{\spaceskip=\fontdimen2\font plus
\BIBentryALTinterwordstretchfactor\fontdimen3\font minus
  \fontdimen4\font\relax}
\providecommand{\BIBforeignlanguage}[2]{{%
\expandafter\ifx\csname l@#1\endcsname\relax
\typeout{** WARNING: IEEEtran.bst: No hyphenation pattern has been}%
\typeout{** loaded for the language `#1'. Using the pattern for}%
\typeout{** the default language instead.}%
\else
\language=\csname l@#1\endcsname
\fi
#2}}
\providecommand{\BIBdecl}{\relax}
\BIBdecl

\bibitem{Funaki_1}
T.~Funaki and K.~Matsuura, ``Feasibility of the low frequency {AC}
  transmission,'' in \emph{Power Engineering Society Winter Meeting, 2000.
  IEEE}, vol.~4, 2000, pp. 2693--2698 vol.4.

\bibitem{Fischer_1}
W.~Fischer, R.~Braun, and I.~Erlich, ``Low frequency high voltage offshore grid
  for transmission of renewable power,'' in \emph{2012 3rd IEEE PES Innovative
  Smart Grid Technologies Europe (ISGT Europe)}, Oct 2012, pp. 1--6.

\bibitem{PSERC_1}
``Low frequency transmission,'' PSERC Publication, Tech. Rep., 2012.

\bibitem{Tuan_1}
T.~Ngo, M.~Lwin, and S.~Santoso, ``Steady-state analysis and performance of low
  frequency ac transmission lines,'' \emph{IEEE Transactions on Power Systems},
  vol.~31, no.~5, pp. 3873--3880, Sept 2016.

\bibitem{Tuan_2}
T.~Ngo, Q.~Nguyen, and S.~Santoso, ``Voltage stability of low frequency ac
  transmission systems,'' in \emph{2016 IEEE/PES Transmission and Distribution
  Conference and Exposition (TnD)}, May 2016, pp. 1--5.

\bibitem{Rosewater_1}
D.~Rosewater, Q.~Nguyen, and S.~Santoso, ``Optimal field voltage and energy
  storage control for stabilizing synchronous generators on flexible ac
  transmission systems,'' in \emph{2018 IEEE/PES Transmission and Distribution
  Conference and Exposition (T D)}, April 2018, pp. 1--9.

\bibitem{Tom_1}
T.~Russell, ``{ScottishPower Renewables Investigates low frequency transmission
  for {EA3}},'' \url
  {https://www.4coffshore.com/windfarms/scottishpower-renewables-investigates-low-frequency-transmission-for-ea3-nid4661.html},
  Tech. Rep., 2016.

\bibitem{Erlich_1}
I.~Erlich, F.~Shewarega, H.~Wrede, and W.~Fischer, ``Low frequency ac for
  offshore wind power transmission - prospects and challenges,'' in \emph{11th
  IET International Conference on AC and DC Power Transmission}, Feb 2015, pp.
  1--7.

\bibitem{Ruddy_2}
J.~Ruddy, R.~Meere, C.~O’Loughlin, and T.~O’Donnell, ``Design of vsc
  connected low frequency ac offshore transmission with long hvac cables,''
  \emph{IEEE Transactions on Power Delivery}, vol.~33, no.~2, pp. 960--970,
  April 2018.

\bibitem{Quan_3}
Q.~{Nguyen}, G.~{Todeschini}, and S.~{Santoso}, ``Power flow in a
  multi-frequency hvac and hvdc system: Formulation, solution, and
  validation,'' \emph{IEEE Transactions on Power Systems}, vol.~34, no.~4, pp.
  2487--2497, July 2019.

\bibitem{Jef_2}
J.~Beerten, S.~Cole, and R.~Belmans, ``Generalized steady-state {VSC MTDC}
  model for sequential {AC/DC} power flow algorithms,'' \emph{IEEE Transactions
  on Power Systems}, vol.~27, no.~2, pp. 821--829, May 2012.

\bibitem{Baradar_1}
M.~Baradar and M.~Ghandhari, ``A multi-option unified power flow approach for
  hybrid {AC}/{DC} grids incorporating multi-terminal {VSC}-{HVDC},''
  \emph{IEEE Transactions on Power Systems}, vol.~28, no.~3, Aug 2013.

\bibitem{Quan_0}
Q.~{Nguyen}, T.~{Ngo}, and S.~{Santoso}, ``Power flow solution for
  multi-frequency {AC} and multi-terminal {HVDC} power systems,'' in \emph{2016
  IEEE Power and Energy Society General Meeting (PESGM)}, July 2016, pp. 1--5.

\bibitem{Farivar_2}
M.~Farivar and S.~H. Low, ``Branch flow model: Relaxations and
  convexification—part i,'' \emph{IEEE Transactions on Power Systems},
  vol.~28, no.~3, pp. 2554--2564, Aug 2013.

\bibitem{Lavaei_1}
J.~Lavaei and S.~H. Low, ``Zero duality gap in optimal power flow problem,''
  \emph{IEEE Transactions on Power Systems}, vol.~27, no.~1, pp. 92--107, Feb
  2012.

\bibitem{Capitanescu_1}
F.~Capitanescu, ``\BIBforeignlanguage{English}{Critical review of recent
  advances and further developments needed in ac optimal power flow},''
  \emph{\BIBforeignlanguage{English}{Electric Power Systems Research}}, vol.
  136, 2016.

\bibitem{Torres_1}
G.~L. Torres and V.~H. Quintana, ``An interior-point method for nonlinear
  optimal power flow using voltage rectangular coordinates,'' \emph{IEEE
  Transactions on Power Systems}, vol.~13, no.~4, Nov 1998.

\bibitem{Liu_1}
M.~Liu, S.~K. Tso, and Y.~Cheng, ``An extended nonlinear primal-dual
  interior-point algorithm for reactive-power optimization of large-scale power
  systems with discrete control variables,'' \emph{IEEE Transactions on Power
  Systems}, vol.~17, no.~4, pp. 982--991, Nov 2002.

\bibitem{Quan_2}
Q.~{Nguyen}, H.~V. {Padullaparti}, K.~{Lao}, S.~{Santoso}, X.~{Ke}, and
  N.~{Samaan}, ``Exact optimal power dispatch in unbalanced distribution
  systems with high pv penetration,'' \emph{IEEE Transactions on Power
  Systems}, vol.~34, no.~1, pp. 718--728, Jan 2019.

\bibitem{IPOPT}
\BIBentryALTinterwordspacing
A.~W{\"a}chter and L.~T. Biegler, ``On the implementation of an interior-point
  filter line-search algorithm for large-scale nonlinear programming,''
  \emph{Mathematical Programming}, vol. 106, no.~1, pp. 25--57, 2006. [Online].
  Available: \url{http://dx.doi.org/10.1007/s10107-004-0559-y}
\BIBentrySTDinterwordspacing

\bibitem{KNITRO}
\BIBentryALTinterwordspacing
R.~H. Byrd, J.~Nocedal, and R.~A. Waltz, \emph{Knitro: An Integrated Package
  for Nonlinear Optimization}.\hskip 1em plus 0.5em minus 0.4em\relax Boston,
  MA: Springer US, 2006, pp. 35--59. [Online]. Available:
  \url{https://doi.org/10.1007/0-387-30065-1_4}
\BIBentrySTDinterwordspacing

\bibitem{BONMIN}
\BIBentryALTinterwordspacing
P.~Bonami, L.~T. Biegler, A.~R. Conn, G.~Cornuéjols, I.~E. Grossmann, C.~D.
  Laird, J.~Lee, A.~Lodi, F.~Margot, N.~Sawaya, and A.~Wächter, ``An
  algorithmic framework for convex mixed integer nonlinear programs,''
  \emph{Discrete Optimization}, vol.~5, no.~2, pp. 186 -- 204, 2008, in Memory
  of George B. Dantzig. [Online]. Available:
  \url{http://www.sciencedirect.com/science/article/pii/S1572528607000448}
\BIBentrySTDinterwordspacing

\bibitem{Teodorescu_1}
R.~Teodorescu, M.~Liserre, and P.~Rodrguez, \emph{Grid converters for
  photovoltaic and wind power systems}.\hskip 1em plus 0.5em minus 0.4em\relax
  John Wiley, Ltd, 2011.

\bibitem{Li_1}
S.~Li, T.~A. Haskew, and L.~Xu, ``Control of {HVDC} light system using
  conventional and direct current vector control approaches,'' \emph{IEEE
  Transactions on Power Electronics}, vol.~25, no.~12, Dec 2010.

\bibitem{Jef_1}
J.~Beerten, R.~Belmans, and S.~Cole, ``A sequential {AC/DC} power flow
  algorithm for networks containing multi-terminal {VSC HVDC} systems,'' in
  \emph{IEEE PES General Meeting}, July 2010, pp. 1--7.

\bibitem{Feng_1}
W.~Feng, L.~A. Tuan, L.~B. Tjernberg, A.~Mannikoff, and A.~Bergman, ``A new
  approach for benefit evaluation of multiterminal {VSC HVDC} using a proposed
  mixed {AC-DC} optimal power flow,'' \emph{IEEE Transactions on Power
  Delivery}, vol.~29, no.~1, pp. 432--443, Feb 2014.

\bibitem{ABB_1}
ABB, ``{It is time to connect with offshore wind supplement},'' \url
  {https://library.e.abb.com/public/1bad1970cd0766eec1257b28005757df/Pow0038%20R6%20LR.pdf},
  Tech. Rep., 2010.

\bibitem{TestTransmissionSystems}
\BIBentryALTinterwordspacing
Literature-based power flow test cases. [Online]. Available:
  \url{{http://icseg.iti.illinois.edu/power-cases/}}
\BIBentrySTDinterwordspacing

\end{thebibliography}
\bibliographystyle{IEEEtran}

\begin{IEEEbiographynophoto}{Quan Nguyen}
	(S'15) is currently pursuing a PhD at The University of Texas at Austin. His research interests are power system control and optimization, renewable energy integration, power quality, and power electronics.
\end{IEEEbiographynophoto}
%\begin{IEEEbiographynophoto}{Quan Nguyen}
%	(S'15) received his Bachelor’s degree from Hanoi University of Science and Technology, Vietnam in 2012 and the M.S degree from The University of Texas at Austin, USA in 2016, both in Electrical Engineering. He is currently pursuing a PhD at The University of Texas at Austin. His research interests are power system modeling and
%	optimization, renewable energy integration, power quality, and power electronics.
%\end{IEEEbiographynophoto}

\begin{IEEEbiographynophoto}{Keng-Weng Lao}
	(S'09-M'17) is a Lecturer in the Department of Electrical and Computer Engineering, University of Macau.  His current research interests include FACTS compensation devices, renewable energy, energy saving, and energy management.
\end{IEEEbiographynophoto}

\begin{IEEEbiographynophoto}{Phuong Vu}
	is a Lecturer of the School of Electrical Engineering, Hanoi University of Science and Technology. His research interests include power electronics and control, electrical machine drive, and renewable energy integration.
\end{IEEEbiographynophoto}

\begin{IEEEbiographynophoto}{Surya Santoso}
	(F'15) is Professor of Electrical and Computer Engineering at the University of Texas at Austin. His research interests include power quality, power systems, and renewable energy integration in transmission and distribution systems.  He is co-author of Electrical Power Systems Quality (3rd edition), sole author of Fundamentals of Electric Power Quality, and editor of Handbook of Electric Power Calculations (4th edition) and Standard Handbook for Electrical Engineers (17th edition).  He is an IEEE Fellow.
	
\end{IEEEbiographynophoto}

\end{document}